\documentclass[twocolumn,english,american,aps,pra,floatfix,eqsecnum]{revtex4-1}
\usepackage[T1]{fontenc}
\usepackage[latin9]{inputenc}
\usepackage{bm}
\usepackage{amsbsy}
\usepackage{graphicx}
\usepackage{esint}

\makeatletter
\@ifundefined{date}{}{\date{}}
\makeatother

\usepackage{babel}
\begin{document}

\title{Simulating and assessing boson sampling experiments with phase-space
representations}

\author{Bogdan~Opanchuk$^{1}$, Laura~Rosales-Z\'arate$^{1,2}$, Margaret~D~Reid$^{1}$
and Peter~D~Drummond$^{1,3}$}

\affiliation{$^{1}$ Centre for Quantum and Optical Science, Swinburne University
of Technology, Melbourne 3122, Australia}

\affiliation{$^{2}$Centro de Investigaciones en \'Optica A.C., Le\'on, Guanajuato
37150, M\'exico}

\affiliation{$^{3}$ Kavli Institute for Theoretical Physics, UC Santa Barbara,
USA}
\begin{abstract}
The search for new, application-specific quantum computers designed
to outperform any classical computer is driven by the ending of Moore's
law and the quantum advantages potentially obtainable. Photonic networks
are promising examples, with experimental demonstrations and potential
for obtaining a quantum computer to solve problems believed classically
impossible. This introduces a challenge: how does one design or understand
such photonic networks? One must be able to calculate observables
using general methods capable of treating arbitrary inputs, dissipation
and noise. We develop novel complex phase-space software for simulating
these photonic networks, and apply this to boson sampling experiments.
Our techniques give sampling errors orders of magnitude lower than
experimental correlation measurements for the same number of samples.
We show that these techniques remove systematic errors in previous
algorithms for estimating correlations, with large improvements in
errors in some cases. In addition, we obtain a scalable channel-combination
strategy for assessment of boson sampling devices.
\end{abstract}
\maketitle

\section{Introduction}

Bosonic quantum systems have an exponential complexity which has long
been recognized as a fundamental theoretical challenge. At the same
time, such complexity is also a potential resource in quantum technology.
One solution to this challenge is in the application of coherence
theory to these problems. Yet very little of the intuition and power
of coherence theory\cite{Glauber1963_CoherentStates} and quasi-probability
theory\cite{Glauber_1963_P-Rep} has been brought to bear on the problems
of quantum technology. As an example, the quantum statistical properties
of bosonic networks are almost exclusively discussed in terms of number
state representations. Such treatments can be useful, but quickly
run into the inevitable complexity limits of number state expansions,
when applied to networks of large size.

Linear bosonic networks are now being used to implement novel quantum
technologies, including boson sampling~\cite{AaronsonArkhipov:2011,AaronsonArkhipov2013LV}
and high accuracy, quantum Fourier interferometers~\cite{Motes2015_PRL114}.
In this paper, we utilize complex P-representation methods~\cite{Drummond_Gardiner_PositivePRep}
that can treat any quantum inputs and outputs to such networks, even
at large sizes. The novelty of our approach is in a transformation
from an exponentially hard number state problem, to a simpler one
in which coherent states with equivalent moments and correlations
is sampled. These techniques help analyze quantum hardware. As an
application, we treat the open problem of how to \emph{assess} that
boson sampling experiments work correctly. We propose a strategy for
assessment that is scalable to a large network size. 

In such experiments one prepares an $M$-mode bosonic state $\hat{\rho}$,
which is input into a passive linear optical multimode device, followed
by a measurement on the output~\cite{Carolan2015}. For $N$ single
boson inputs into multiple channels, the generation of the random
output counts is an exponentially hard computational problem~\cite{AaronsonArkhipov2013LV,Aaronson2011}.
The most advanced known classical algorithm~\cite{Clifford2017-classical}
takes a time exponential in \emph{N} to generate a single random sample
of this type, which makes it impractical at large $N$. Quantum Fourier
interferometers are an application of this quantum technology in high-precision
metrology~\cite{Su:2017}. However, most of the development of the
theory has taken place using orthogonal number states, which have
inherent limits when treating large numbers of bosons or modes. Our
methods are based on general quantum phase-space representation theorems,
so that the underlying techniques are generally applicable to linear
photonic networks.

Boson sampling experiments~\cite{Motes2015_PRL114,Carolan2015,Broome2013,Crespi2013,Tillmann2013,Spring72013,Crespi:2014,Spagnolo2014,Wang2017}
have the goal of demonstrating computations thought to be impossible
on classical computers, and lay the foundations for new quantum technologies.
To fully realize this potential, one must have tools to analyze them.
Ultimately it will be necessary to take account of imperfect inputs,
as well as losses and other non-ideal behavior. Here, we derive a
hybrid computational and analytic approach to allow the assessment
of such networks. This algorithm is not intended to solve the BosonSampling
problem of generating the random counts. However, it does make it
feasible to analyze, design and assess photonic networks used for
this and other quantum technology applications.

\begin{figure}
\includegraphics{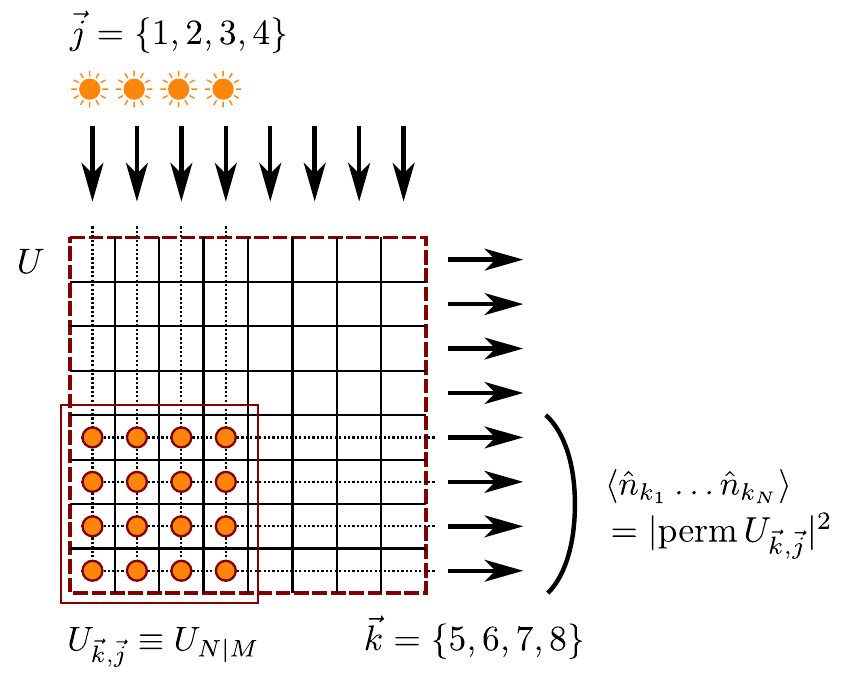}\caption{\label{fig:Channel-diagram-of}Channel diagram of a boson sampling
photonic network gedanken experiment.}
\end{figure}

Linear photonic networks are defined by an $M\times M$ mode unitary
matrix $\bm{U}$, or more generally by an input-output transformation
matrix $\bm{T}$, which can include losses. The simplest boson sampling
experiments~\cite{Broome2013,Crespi2013,Tillmann2013,Spring72013,Crespi:2014,Loredo2016,Spagnolo2014}
have an initial $N$-photon state $\left|\bm{n}\right\rangle $. Here
$\bm{n}$ is a vector such that $N=\sum n_{j}$, $n_{j}=0,1$, and
$\left|n\right\rangle $ is the number state basis. More generally,
one may anticipate that other nonclassical input states will be used.
The output photon numbers, $\bm{n}^{\prime}$, are the observables.
The permanent-squared of the sub-matrices defined by the input and
output modes~\cite{Scheel2004Permanents,Scheel2005} gives the probability
of measuring one photon in each preselected output mode~\cite{Motes2015_PRL114,Tillmann2013},
as illustrated in Fig.~\ref{fig:Channel-diagram-of}. The optimal
classical techniques for calculating this $\#P$ hard problem~\cite{Valiant1979}
scale exponentially as $N2^{N}$~\cite{Glynn2010Perm} operations
for $N$ inputs, and become rapidly infeasible above $N=50$~\cite{Wu2016arXiv160605836}. 

To understand the reason for this scaling, we note that the permanent
of a square matrix $\bm{U}$ is defined as a sum over all permutations
$\boldsymbol{m}$ of the set of indices $\{1,\dots,N\}$, ${\rm perm}(\bm{U})=\sum_{\boldsymbol{m}}\prod_{j}U_{jm_{j}}$.
The number of such permutations is $N!$. This complexity occurs naturally
in a photonic network, driven by the enormous number of possible interference
paths that a photon can take. Typically, the filled channels $N$
are only a small fraction of the open channels $M$, so that $N=M/k$
where $k\gg1$. Thus, not only is the calculation of each permanent
exponentially hard as $N$ increases, but in addition there are exponentially
many possible sets of $N$ output channels. Each of these combinations
corresponds to a different, unique permanent~\textemdash{} all exponentially
hard~\textemdash{} which must be calculated to predict the full set
of output probabilities. While it is known~\cite{Clifford2017-classical}
that there exists a random bit generation algorithm that is only exponential
in $N$ (rather than $M$), for generating single random samples,
this does not solve the assessment problem.

We simulate such linear photonic experiments using ``quantum simulation
software'', by transforming this problem into an exact expansion
on non-classical phase space~\cite{Drummond_Gardiner_PositivePRep},
combined with a random sampling method that generates single phase-space
samples with an algorithm that is only polynomial in $N$ and $M$.
The approach allows unbiased estimates of the experimentally measured
probabilities, and scales \emph{better} than an experiment, in terms
of time taken for a correlation measurement with a given sampling
error. This is an important point, since theories used to assess the
experiments must be calculable in a time comparable to the measurements
themselves, in order to be useful. In addition, it removes systematic
errors that can occur when estimating correlations using previous
algorithms. We expect that the general approach will be applicable
to experiments using a range of nonclassical input states.

However, neither measurements nor calculations of correlations can
be readily scaled to large sizes for assessment in a finite time with
typical unitaries. Instead, we propose an analytic signature for assessing
claims to solve Boson Sampling, using a hierarchy of measurable combinations
of $N$-th order correlations, with successive channels deleted. 

Assessment is an integral part of analyzing solutions to exponential
complexity. Our proposal involves sums over exponentially large numbers
of sub-permanents present in the experiments. This is essential to
obtain significant experimental counts. Any assessment is limited
by experimental and theoretical scaling limits, due to count rate
issues. Our analytic theory itself involves a conjectured solution
that is verified numerically. It is likely that it can be rigorously
proved using random matrix theory. 

Finally, our approach can calculate in principle any input state or
output measurement. More details of these applications to quantum
Fourier transform interferometry including dissipative effects from
phase noise can be found elsewhere~\cite{Opanchuk2017-robustness}.
This is essential for understanding how this technology can be scaled
to large sizes with imperfect sources, with recent progress in this
direction for boson sampling using time-binning techniques~\cite{He2016ScalableBS}
and in metrology applications~\cite{OlsonLinearOptQM}. 

The contents of the paper are as follows. Section~\ref{sec:Complex-P-representation}
explains how the complex-P continuous phase-space representation can
be used to calculate arbitrary observables given any quantum input
state to a photonic network, while Section~\ref{sec:Discrete-complex-P}
gives a discrete version of this method. Section~\ref{sec:Methods-for-sampling}
gives sampled results for both these approaches, using a randomized
calculation of a boson sampling experiment. This section includes
a treatment of scaling behavior and a comparison both to experimental
sampling errors and to the Gurvits approximation for permanents. Section~\ref{sec:combined-correlations}
explains how these results are generalized to obtain a scalable assessment
method for boson sampling, with comparisons to some other proposals.
Section~\ref{sec:Conclusion} gives our conclusions.

\section{Complex P-representation\label{sec:Complex-P-representation}}

We first explain our phase-space method, which uses the complex P-distribution.
This is a quantum phase-space expansion over a basis of coherent state
projectors~\cite{Drummond_Gardiner_PositivePRep}. It is an extension
of methods developed originally by Glauber~\cite{Glauber_1963_P-Rep}
in quantum optics, and has the capability of treating any input quantum
density matrix. We use this method to simulate linear photonic experiments,
evaluating the measurable photonic moments by probabilistically sampling
over specifically selected contours in the higher dimensional complex
space $\mathcal{C}^{M}\times\mathcal{C}^{M}$. Our simulations \emph{do
not }generate a classical sequence of photon counts with a permanental
distribution, which is known to be exponentially hard. Instead we
employ something more useful for assessment purposes: a sequence of
numbers that generates equivalent correlations and moments to an experiment,
but has much \emph{lower} sampling errors for the same number of samples
used. 

The integration contour is illustrated schematically in Fig.~\ref{fig:Contour integral},
noting that only one complex variable of the $2N$ that are integrated
is shown. We note that this simulation does not directly model photon-counting
measurements, since there is a different representation for every
operator ordering and measurement, as originally mentioned in Dirac's
review~\cite{Dirac_RevModPhys_1945}. The advantage of this approach
is that the simulation is no longer exponentially hard to carry out
and can still be used to evaluate correlations or moments of photon-counts,
which has important applications for assessment protocols.

\subsection{Contour integrals and input states}

\begin{figure}
\includegraphics{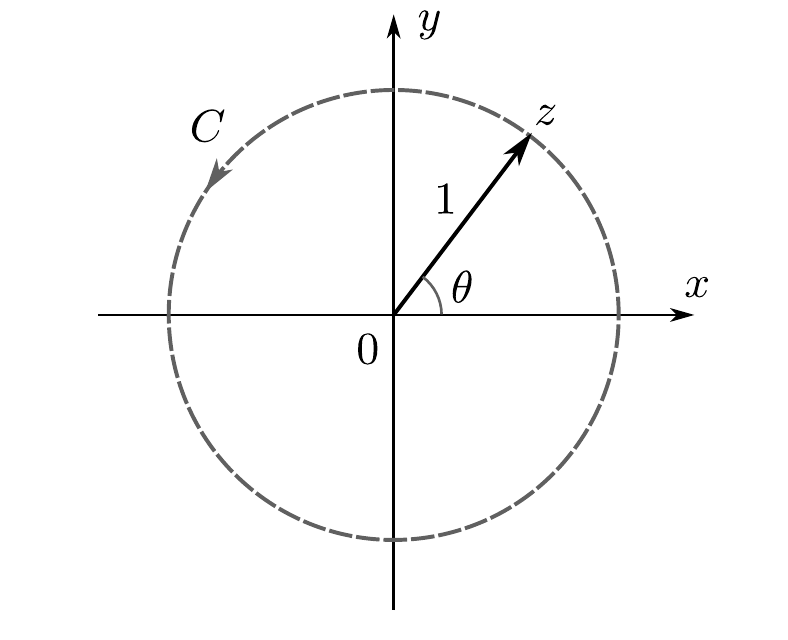}\caption{\label{fig:Contour integral}Contour integral for $z$, where $\alpha=rz$.
Here $z$ is chosen randomly with unit modulus for sampling purposes,
to give a Monte-Carlo sampled contour integral.}
\end{figure}

There are several different generalized P-representations, including
complex and positive valued representations~\cite{Drummond_Gardiner_PositivePRep}.
Because it has a compact form, without large radius tails, the complex
P-representation is very scalable when treating the high-order correlations
measured in boson sampling experiments. With this representation,
the input quantum density matrix $\hat{\rho}^{(\mathrm{in})}$ is
represented by an integral over a closed contour $C$ enclosing the
origin in a multi-dimensional complex plane as in Fig.~\ref{fig:Contour integral}:
\begin{equation}
\hat{\rho}^{(\mathrm{in})}=\oiint_{C}P(\bm{\alpha},\bm{\beta})\hat{\Lambda}\left(\bm{\alpha},\bm{\beta}\right)\mathrm{d}\bm{\alpha}\mathrm{d}\bm{\beta}\,.\label{eq:Complex-P}
\end{equation}
 Here, $P(\bm{\alpha},\bm{\beta})$ is a complex distribution function
that is a function of the input photon numbers, while $\bm{\alpha}=\left[\alpha_{1},\cdots\alpha_{M}\right]$
is a vector of $M$ complex numbers, as is $\bm{\beta}$. The quantum
operator basis $\hat{\Lambda}$ is a set of generalized coherent state
projectors:
\begin{equation}
\hat{\Lambda}\left(\bm{\alpha},\bm{\beta}\right)=\frac{\left\Vert \bm{\alpha}\right\rangle \left\langle \bm{\beta}^{*}\right\Vert }{\left\langle \bm{\beta}^{*}\right\Vert \left.\bm{\alpha}\right\rangle }\,,
\end{equation}
onto un-normalized Bargmann-Glauber~\cite{Glauber1963_CoherentStates}
coherent states $\left\Vert \bm{\alpha}\right\rangle $. These are
defined using photon number states $\left|n_{k}\right\rangle $ with
$n_{k}$ photons in the $k$-th input mode,
\begin{equation}
\left\Vert \bm{\alpha}\right\rangle \equiv\prod_{k}\left[\sum_{n_{k}=0}^{\infty}\frac{\alpha_{k}^{n_{k}}}{\sqrt{n_{k}!}}\left|n_{k}\right\rangle \right]\,.
\end{equation}
For the purposes of the sampling, it is only necessary to know that
$\alpha_{k}$ and $\beta_{k}$ are complex numbers, and that the phase
space method will generate the correct quantum moments. We integrate
around a circular contour of radius $r$, by randomly sampling unit
modulus complex numbers $z,z'$, where $\alpha=rz$ and $\beta=rz'$
, with a complex-valued weight $P\left(\bm{\alpha},\bm{\beta}\right)$.
The contour is illustrated in Fig.~\ref{fig:Contour integral}.

The over-completeness of the coherent states in quantum mechanics
means that there is more than one way to choose the contour $C$,
and in particular the circular radius $r$ can be varied. It is an
essential feature of our method that this flexibility enables us to
tailor the representation to optimize the sampling for different tasks.

For greater accuracy when developing sampling techniques to be used
later, one can also use the fact that the density matrix is hermitian.
In this refinement, we define the expansion in an explicitly hermitian
form by taking the real part, thus imposing the constraint that the
final result must be real:
\begin{equation}
\hat{\rho}^{(\mathrm{in})}=\Re\oiint_{C}P(\bm{\alpha},\bm{\beta})\hat{\Lambda}\left(\bm{\alpha},\bm{\beta}\right)\mathrm{d}\bm{\alpha}\mathrm{d}\bm{\beta}\,.\label{eq:Complex-P-1}
\end{equation}

To perform simulations using this approach it is necessary to have
a representation, $P\left(\bm{\alpha},\bm{\beta}\right)$, of the
initial state. This distribution exists for an \emph{arbitrary} initial
density matrix, either pure or mixed, with any coherence properties
and photon numbers~\cite{Drummond_Gardiner_PositivePRep}. In this
paper, we focus on the simplest case: a pure state in which the $k$-th
channel has an $n_{k}$-boson input. The usual boson sampling experiments
send one photon into each of a set $\sigma$ of $N$ input modes,
so that $n_{k}=0,1$, and, therefore:
\[
N=\sum_{k}n_{k}=\sum_{k\in\sigma}n_{k}\,.
\]
There is more than one way to represent these states, which are illustrated
below. However, the same techniques can be readily adapted to treat
other quantum inputs, for example squeezed or entangled states, which
can be more easily generated in experiments.

\subsection{Output measurements}

Any bosonic input density matrix $\hat{\rho}^{(\mathrm{in})}$ is
changed by transmission through a linear network to an output density
matrix $\hat{\rho}^{(\mathrm{out})}$. In general, to take into account
inevitable losses, one can imagine a larger unitary that includes
loss channels, but only consider the sub-matrix $\bm{T}$ for the
accessible channels that are measured, with other channels just being
ignored. An example is the transmission matrix $\bm{T}=\sqrt{t}\bm{U}$
in which all channels experience equal loss. This combines the $M\times M$
unitary mode transformation $\bm{U}$ of the entire network, with
an absorptive transmission coefficient $t$. 

The effect of the transmission matrix $\bm{T}$ on the phase-space
distribution is straightforward, owing to the normal ordering property
of the complex P-representation. It simply transforms the coherent
amplitudes in a deterministic way~\cite{drummond2014quantum}, such
that $\bm{\alpha}^{(\mathrm{out})}=\bm{T}\bm{\alpha},\,\bm{\beta}^{(\mathrm{out})}=\bm{T}^{*}\bm{\beta}$.
The resulting output density matrix is therefore a contour integral
with the same weight, but a modified projector:
\begin{equation}
\hat{\rho}^{(\mathrm{out})}=\Re\oiint_{C}P(\bm{\alpha},\bm{\beta})\hat{\Lambda}\left(\bm{T}\bm{\alpha},\,\bm{T}^{*}\bm{\beta}\right)\mathrm{d}\bm{\alpha}\mathrm{d}\bm{\beta}\,.\label{eq:Output-P}
\end{equation}

Any output number correlation is given by computing moments of the
known input $P$-function, using the equivalent output number variable,
\begin{eqnarray}
n_{k}^{(\mathrm{out})}(\bm{\alpha},\bm{\beta}) & = & \alpha_{k}^{(\mathrm{out})}\beta_{k}^{(\mathrm{out})}\\
 & = & \left(\sum_{j}T_{kj}\alpha_{j}\right)\left(\sum_{j}T_{kj}^{*}\beta_{j}\right)\,.\nonumber 
\end{eqnarray}
A typical observable in these experiments~\cite{Tichy2014,Mayer2011,Walschaers2016_NJP18}
is an arbitrary normally ordered quantum correlation of $k$-th mode
operators, $\hat{n}_{k}\equiv\hat{a}_{k}^{\dagger}\hat{a}_{k}$, belonging
to a set of output modes $\sigma^{\prime}$. In boson sampling we
are interested in the correlator over $N$ output channels. Since
it is assumed that no photons are generated inside the device~\textemdash{}
it has no gain~\textemdash{} any $N$-fold coincidence count means
that there is exactly one photon detected in each output channel.
The correlator or coincidence count therefore has accessible eigenvalues
of $1,0$.

This property of binary counts means that the correlator is also the
probability of an $N$-fold coincidence count, given an $M\times M$
transmission matrix,
\begin{equation}
P_{N|M}\equiv\left\langle \prod_{k\in\sigma'}\hat{n}_{k}\right\rangle _{\mathrm{Q}}\,,
\end{equation}
where quantum expectation values are denoted $\left\langle \right\rangle _{Q}$.
In other words, the correlation and the count probability are identical.
Since these observables are hermitian, we can take the real part of
the expansion as in Eq.~(\ref{eq:Output-P}). The quantity of interest
here is therefore given by the following expression for $P_{N|M}$,
where a contour integral over the complex P-function is denoted as
$\left\langle \right\rangle _{P}$:
\begin{eqnarray}
P_{N|M} & = & \Re\oiint_{C}P(\bm{\alpha},\bm{\beta})\prod_{k\in\sigma^{\prime}}n_{k}^{(\mathrm{out})}(\bm{\alpha},\bm{\beta})\mathrm{d}\bm{\alpha}\mathrm{d}\bm{\beta}\nonumber \\
 & \equiv & \Re\left\langle \prod_{k\in\sigma'}n_{k}^{(\mathrm{out})}(\bm{\alpha},\bm{\beta})\right\rangle _{\mathrm{P}}.\label{eq:Outputs}
\end{eqnarray}

As the representation $P$ is not unique (the choice of contour being
flexible), one can choose different representations of the input state.
The main point that we wish to emphasize is that these techniques
comprise a \emph{complete} set of efficient strategies for treating
quantum photonic networks. Even though we focus here on number state
inputs and outputs, the general approach is not limited to this case.
Such methods have previously been used in nonlinear quantum problems
as well, and give a general strategy for calculating any normal-ordered
observable correlation~\cite{Glauber1963_CoherentStates}.

The complex P-representation is known to have exact solutions for
the non-equilibrium steady-state solutions in other quantum photonic
devices. These include the driven anharmonic cavity~\cite{drummond1980quantum,Bartolo:2016,Minganti:2016},
the nonlinear two-photon absorber~\cite{DrummondGardinerWalls1981},
the degenerate parametric oscillator~\cite{drummond1981_II_nonequilibriumparamp},
the non-degenerate parametric oscillator~\cite{mcneil1983quantum}
as well as approximate solutions for coupled nonlinear cavities~\cite{Cao:2016}
and Bose condensates~\cite{drummond2016coherent}. Thus, it may not
be impossible to extend these techniques to nonlinear photonic networks.

\section{Discrete complex P-representation\label{sec:Discrete-complex-P}}

Just as the coherent states do not provide unique expansions of quantum
states, neither do complex P-representations. This provides an opportunity
to adapt the representation to a given task. In this section, we recall
the most well-known type of complex P-representation, and also explain
an alternative form using discrete sums, which is more suitable to
sampling high-order correlations. For number state inputs, the complex
P-distribution has a well-known solution that is readily proved to
exist using Cauchy's theorem:
\begin{equation}
P\left(\bm{\alpha},\bm{\beta}\right)=\prod_{k}\left[\left(\frac{n_{k}!}{2\pi\mathrm{i}}\right)^{2}\frac{e^{\alpha_{k}\beta_{k}}}{(\alpha_{k}\beta_{k})^{n_{k}+1}}\right]\,,\label{eq:Complex-P-Dist}
\end{equation}
where $n_{k}$ is the photon count in the mode $k$. For $n_{k}=0$,
the single pole at the origin means that one can replace the input
variable by its vacuum value of $\alpha_{k}=\beta_{k}=0$ in any average,
so that $P\left(\alpha_{k},\beta_{k}\right)=\delta\left(\alpha_{k}\right)\delta\left(\beta_{k}\right)$,
where for the case of a vacuum mode input into a single mode $k$.

\subsection{Discrete summation method}

While the contour integral solution given above always exists, other
forms are also possible. These give a different strategy allowing
efficient random sampling methods. Here we will use the limit of an
infinitely small contour, which allows us to use a discrete phase
summation method. Other choices can be utilized for more general states.
Discrete phase sums have particular utility in cases where the initial
photon number is bounded, for example with a fixed input boson number.
We now introduce one of these discrete approaches, which we term the
discrete or qudit complex P-representation (QCP)~\cite{Drummond2016-coherent-states}.
This construction is useful in the limit of the circular radius $r\rightarrow0$,
which leads to $d$ coherent phases distributed on an infinitesimal
circle. 

This alternative solution for $P\left(\bm{\alpha},\bm{\beta}\right)$,
is still in the form of Eq.~(\ref{eq:Complex-P}), but with the distribution
consisting of $d$ discrete delta-functions arranged on a circle of
radius $r\rightarrow0$. With appropriate choices of the complex amplitude
at each point, one can represent an arbitrary quantum state of a $d$-dimensional
qudit, with initial photon occupation numbers in each mode of up to
$n=d-1$. We will show that this approach unifies quantum representation
theory~\cite{Drummond_Gardiner_PositivePRep} with discrete sampling
permanent approximation methods~\cite{Gurvits2005}. It has numerical
properties that make it both accurate and efficient. 

In this approach the input quantum state is expanded as a superposition
of a discrete set of coherent amplitudes in each of the $N$ non-vacuum
modes, defined as:
\begin{eqnarray}
\alpha^{(q)} & = & rz^{q}\quad q=0,\ldots d-1\,\\
\beta^{(\tilde{q})} & = & rz^{-\tilde{q}},\quad\tilde{q}=0,\ldots d-1\,,\nonumber 
\end{eqnarray}
where $z=\exp\left(i\phi\right)$ is the $d$-th root of unity and
the phase interval is $\phi=2\pi/d$. The density matrix is then expanded
in coherent states with a discrete summation for the $N$ modes that
are not initially in the vacuum state, using a discretized complex
P-function, $P_{Q}(\boldsymbol{q},\tilde{\boldsymbol{q}})$ which
is expanded in terms of a number-projected kernel, $\hat{\Lambda}^{(d)}$:
\begin{equation}
\hat{\rho}=\frac{1}{d^{2M}}\sum_{\boldsymbol{q},\tilde{\boldsymbol{q}}}P_{Q}(\boldsymbol{q},\tilde{\boldsymbol{q}})\hat{\Lambda}^{(d)}(\boldsymbol{q},\tilde{\boldsymbol{q}})\,.
\end{equation}
Here $\hat{\Lambda}^{(d)}(\boldsymbol{q},\tilde{\boldsymbol{q}})=\mathcal{P}\left\Vert \bm{\alpha}^{(\boldsymbol{q})}\right\rangle \left\langle \bm{\beta}^{(\tilde{\boldsymbol{q}})*}\right\Vert $,
where $\mathcal{P}$ is a projector onto a subspace of up to $d-1$
photons per input mode, and the coherent amplitudes are:
\begin{eqnarray}
\bm{\alpha}^{(\boldsymbol{q})} & \equiv & \left[\alpha^{\left(q_{1}\right)},\alpha^{\left(q_{2}\right)},\ldots\right]\nonumber \\
\bm{\beta}^{(\tilde{\boldsymbol{q}})} & \equiv & \left[\beta^{\left(\tilde{q}_{1}\right)},\beta^{\left(\tilde{q}_{2}\right)},\ldots\right].
\end{eqnarray}
 This form of representation, the complex qudit P-distribution (QCP),
can be shown to always exist for Hilbert spaces of bounded occupation
numbers, using discrete Fourier transforms. While the full derivation~\cite{Drummond2016-coherent-states}
is given elsewhere, there is an important point to be noted. The discrete
form is combined with a projection operator at finite $r$. However,
these coherent states have unit norm in the limit of $r\rightarrow0$,
and in this limit no projection is needed. Input states of higher
photon number are excluded automatically, either from Fourier orthogonality
or because they have negligible weight in this limit.

Therefore, using the above expansion, a complex qudit P-function $P_{\mathrm{Q}}$
always exists for the input density matrix $\hat{\rho}$ used here,
where:
\begin{equation}
P_{\mathrm{Q}}(\boldsymbol{q},\tilde{\boldsymbol{q}})=\sum_{\bm{n},\bm{m}}\left\langle \bm{m}\right|\hat{\rho}\left|\bm{n}\right\rangle \prod_{j}\frac{\sqrt{n_{j}!m_{j}!}}{r^{n_{j}+m_{j}}}e^{\left[\mathrm{i}\phi(n_{j}\tilde{q}_{j}-m_{j}q_{j})\right]}\,.\label{eq:General}
\end{equation}

Here the expansion allows occupation numbers for mode $k$ up to $n_{k}=0,\ldots d-1$.
We note that such discrete sampling gives continuous sampling in the
limit of $d\rightarrow\infty$, and in this limit simply reduces to
the earlier contour integral result of~(\ref{eq:Complex-P-Dist})
at small radius. The discrete expansion can be verified as a solution,
by inserting this distribution into the expansion of the density matrix,
and noting that, from the properties of the discrete Fourier transform,
\begin{equation}
\frac{1}{d^{N}}\sum_{\boldsymbol{q}}\mathrm{e}^{\mathrm{i}\bm{q}\cdot\left(\bm{n}-\bm{m}\right)\phi}=\delta_{\bm{n}-\bm{m}}.
\end{equation}
In the simplest binary, or qubit, case where $d=2$ and $n_{k}=0,1$
we consider $q_{k}=\{0,1\}$. This implies that $\alpha_{k}=\pm r$. 

For finite $d$, this method is only applicable to photon number
inputs that are bounded, which is precisely the case in many boson
sampling experiments, such as Quantum Fourier interferometry~\cite{Motes2015_PRL114}.
It is also the case in the assessment scheme described in Section~\ref{sec:combined-correlations}.
Since this is a case of the complex P-representation, the input-output
transformation used previously in~(\ref{eq:Outputs}) is still valid.
The output coherent amplitude for a given discrete input $\bm{\alpha}^{(\boldsymbol{q})}$
is therefore $\bm{T}\bm{\alpha}^{(\boldsymbol{q})}$. This output
is no longer restricted to the same input set of discrete phases,
and hence can include other particle numbers, different to those in
each input channel. Physically this means that interference effects
can occur, leading to the coherent Hong-Ou-Mandel type phenomena that
are responsible for the nonclassical boson sampling output statistics. 

The output photon number phase-space variable, given an input of single
bosons into the first $N$ modes, is:
\begin{equation}
n_{k}^{\mathrm{(out)}}(\bm{\alpha}^{(\bm{q})},\bm{\beta}^{(\tilde{\bm{q}})})=r^{2}\left(\sum_{j\in\sigma}T_{kj}z^{q_{j}}\right)\left(\sum_{j'\in\sigma}T_{kj'}z^{\tilde{q}_{j'}}\right)^{*}.\label{eq:photon-number-variable}
\end{equation}

Here the notation $j\in\sigma$ is used to restrict the sum to the
$N$ occupied input channels. This is then further summed over all
the possible input values of $\bm{q},\tilde{\bm{q}}$, and weighted
with $P_{\mathrm{Q}}(\boldsymbol{q},\tilde{\boldsymbol{q}})$. The
quantum expectation value or probability of observing a simultaneous
count in each of a particular set $\sigma'$ of $N$ output channels
is then given by:
\begin{equation}
P_{N|M}=\frac{1}{d^{2N}}\Re\left[\sum_{\boldsymbol{q},\tilde{\boldsymbol{q}}}P_{Q}(\boldsymbol{q},\tilde{\boldsymbol{q}})\prod_{k\in\sigma'}n_{k}^{(o)}(\bm{q},\tilde{\bm{q}})\right].\label{eq:Exact-coincidence}
\end{equation}

\subsection{P-Representation for single-photon inputs\label{sec:Qudit-sampling}}

We have shown that there is a mapping between the phase-space expansion
and a combinatoric sum. In the simplest one mode, one boson case,
an inspection of the general discrete result, Eq.~(\ref{eq:General}),
shows that the P-function is simply the product of two complex variables
together with a radial factor:
\begin{equation}
P_{\mathrm{Q}}(q,\tilde{q})=\frac{1}{r^{2}}z^{\tilde{q}}z^{-q}\,.\label{eq:single-mode-P-function}
\end{equation}

The advantage of the small radius limit is that the known identities
for the generalized P-representation are all valid, and the $r\rightarrow0$
limit can be taken after the calculation. Thus, we can use the standard
result that after transmission through a linear optical system with
phase-shifts, beam-splitters and losses, the output coherent amplitudes
are multiplied by the relevant linear transmission matrix. In calculating
$N$-th order correlations of an $N$-photon input, all the factors
proportional to the radius $r$ simply cancel. 

As a result, after including the complex P-function weights, an $N$-th
order output correlation is:
\begin{eqnarray}
P_{N|M} & = & \left|\frac{1}{d^{N}}\sum_{\boldsymbol{q}}\left\{ \prod_{\ell\in\sigma}z^{-q_{\ell}}\prod_{k\in\sigma'}\left(\sum_{j\in\sigma}T_{kj}z^{q_{j}}\right)\right\} \right|^{2}.\nonumber \\
 & = & \left|\mathrm{perm}\left[T\left(\sigma',\sigma\right)\right]\right|^{2}.\label{eq:exact permanent-squared}
\end{eqnarray}

As expected~\cite{Aaronson2011}, this is the square of the permanent
of the sub-matrix of $\bm{T}$ with rows in $\sigma^{\prime}$ and
columns in $\sigma$, which we call $\mathbf{T}\left(\sigma',\sigma\right)\equiv\mathbf{M}$.
After summation on the $\boldsymbol{q}$ indices, the only terms that
survive involve products of distinct permutations of the matrix indices,
which is the permanent. However, there are exponentially many terms
involved at large $N$, which requires a sampling technique that we
explain in the next section. 

We finally note that the complexity for exact calculations depends
on the value of the discretization number, $d$. The minimum value
of $d$ for a boson sampling experiment with $1$ photon per input
channel is $d=2$. This gives the least complexity in an exact computation.
It also directly corresponds to the most efficient and well-known
computational technique for computing permanents.

However, this is only a lower bound. One can use any value of $d\ge2$.
In particular, we can also consider the exact continuum path-integral
as well. This simply corresponds to replacing the sums in Eq.~(\ref{eq:exact permanent-squared})
by integrals over $\bm{\phi}=\left[\phi_{1},\ldots\phi_{N}\right]$,
so that:
\begin{equation}
P_{N|M}=\left|\int\frac{d^{N}\bm{\phi}}{\left(2\pi\right)^{N}}\prod_{\ell}e^{-i\phi_{\ell}}\prod_{k}\left(\sum_{j}M_{kj}e^{i\phi_{j}}\right)\right|^{2}.
\end{equation}

The last expression is also the $d\rightarrow\infty$ limit of the
discrete complex P-representation, which is a complete representation.
It can be generalized to treat arbitrary input states, in which case
one should use the full expression of Eq.~(\ref{eq:General}) to
specify the P-distribution. While an integral is more complex numerically
than a discrete sum, we show in the next section that it is extremely
efficient when random sampling methods are employed.

\section{Methods for sampling phase-space\label{sec:Methods-for-sampling}}

While the results of the previous section are exact, they are also
exponentially complex. Next, we will explain the approximate randomized
technique that is utilized to calculate the photon counting coincidence
probabilities in polynomial time. This does not contradict complexity
theory results, which only prohibit polynomial time methods for direct
generation of random photon counts or the exact evaluation of matrix
permanents. Verifying such coincidence probabilities is an important
step in any assessment procedure for boson sampling devices.

For reasons of efficiency, the combinatoric sums can be evaluated
approximately by taking pairs of randomly chosen integer vectors $\left(\boldsymbol{q}^{(j)},\tilde{\boldsymbol{q}}^{(j)}\right)$,
and averaging over samples of these random phases. In a Hilbert space
of dimension $d$, there are $d$ possible random discrete phases.
This can be taken to the limit of $d\rightarrow\infty$ for a continuously
sampled Monte Carlo integral. 

There are similarities between experimental measurements and the use
of sampled quantum simulations. In both cases there is a sampling
error, since one must calculate or measure results for correlations
using a finite number of samples. The time taken is proportional to
the number of samples used. It is therefore crucial to know how the
average sampling error scales with the number of active channels $N$,
which determines the computational time 

We find some important results, as follows: 
\begin{itemize}
\item Random sampling method is scalable even for the \emph{most} complex
exact method, that is, the $d\rightarrow\infty$ limit.
\item Computed correlations have a lower sampling error than in an experimental
measurement.
\item These techniques eliminate a systematic error that occurs using previous
approximate methods.
\end{itemize}

\subsection{Overview of calculations and sampling procedure}

In order to explain the general procedure, we will summarize the steps
involved in analyzing a photonic network. 
\begin{enumerate}
\item Firstly, the input $M$-mode bosonic state $\hat{\rho}_{0}$ must
be known, which can be any density matrix. For bounded photon number,
a discrete complex-P expansion can be utilized, otherwise a continuous
expansion is necessary. The existence theorems~\cite{Drummond_Gardiner_PositivePRep}
are used to choose a contour and obtain the P-function, $P\left(\bm{\alpha},\bm{\beta}\right)=\left|P\right|e^{i\phi}$,
which is a distribution over a contour in $N$ complex dimensions.
\item The $M\times M$ transmission matrix $T$ must be known or measured.
This can include any arbitrary loss in principle. In cases of low-frequency
phase noise~\textemdash{} for example, from $1/f$ noise~\cite{perlmutter1988inverse,shelby1990phase}~\textemdash{}
the refractive index and hence the transmission matrix is a random
variable, since each repetition of the experiment has a different
$T$. Dispersion and noise correlated on time-scales shorter than
pulse durations require a more sophisticated theory~\cite{Carter:1991,Drummond:2001a,drummond2014quantum}.
\item The input contour is sampled with equal probability, $\left|P\right|$,
to give coherent samples $\bm{\alpha}^{(j)},\bm{\beta}^{(j)}$, together
with the relevant complex phase. These are multiplied by the transmission
matrix giving the output coherent samples whose moments, combined
with the phase factor $\phi$, give the expected quantum correlations.
\end{enumerate}
This gives a detailed procedure that would be followed in the most
general case. The present paper treats the simplest case of considering
correlated single-photon outputs in a predefined set of channels,
without dissipation. More general examples are treated elsewhere.

\subsection{Sampled calculations of the permanent squared}

In both the continuous phase and discrete phase approach, we use
independent random samples of $\boldsymbol{z}^{(j)}=\exp\left(i\phi\bm{q}^{(j)}\right)$,
$\tilde{\boldsymbol{z}}^{(j)}=\exp\left(i\phi\tilde{\bm{q}}^{(j)}\right)$.
These amplitudes of unit modulus correspond to scaled versions of
$\bm{\alpha},\bm{\beta}^{*}$ in the original complex P-representation.
For each channel $(i)$ and sample $(j)$, we have $z_{i}^{(j)}=z^{q_{i}^{(j)}}$.
In order to obtain an unbiased estimate of the absolute value squared,
we use an ensemble of size $L$ of \emph{two independent} conjugate
noise vectors, $\boldsymbol{z},\tilde{\boldsymbol{z}}$. 

To explain the notation, we define a polynomial function $p(\mathbf{M},\boldsymbol{z})$
of the $N\times N$ sub-matrix of interest, as:
\begin{equation}
p(\mathbf{M},\boldsymbol{z})=\prod_{\ell=1}^{N}z_{\ell}^{-1}\prod_{k=1}^{N}\left(\sum_{j=1}^{N}M_{kj}z_{j}\right).
\end{equation}
Each approximation of the permanent, which we denote as $p(\mathbf{M},\boldsymbol{z})$,
is a function of the sub-matrix $\mathbf{M}$ and a noise vector $\boldsymbol{z}$. 

Regardless of the discretization parameter $d$, any of these sampling
methods give an unbiased estimate of the modulus-squared of the permanent,
and are applicable to any type of matrix permanent calculation, although
the error scaling depends on the algorithm and type of matrix. We
define $\left\langle \right\rangle _{L}$ as the stochastic expectation
value over an ensemble of $L$ random samples of a stochastic vector.
The sampled calculation of a permanent is then given by:
\begin{equation}
\left\langle p(\mathbf{M},\boldsymbol{z})\right\rangle _{L}\equiv\frac{1}{L}\sum_{j=1}^{L}p(\mathbf{M},\boldsymbol{z}^{(j)})\,.\label{eq:subensemble-average}
\end{equation}
When applied to calculating a permanent, with the discretization value
of $d=2$, this method corresponds to a previously known permanent
approximation method, known as the Gurvits approximation~\cite{Gurvits2005}. 

We do not use this more traditional approach. The methods described
here lead to a much larger class of permanent approximations, as well
as algorithms that are specifically optimized for calculating the
\emph{modulus squared} of the permanent~\textemdash{} which is the
quantity of interest in boson sampling. In fact, the optimal sampling
method for the modulus squared is not simply using the permanent algorithm
and squaring the result. The drawback with this method is that when
taking the modulus squared of the stochastic estimate, an additional
term is obtained equal to the sampling variance. Hence, the resulting
estimate is not unbiased.

Instead, following the derivation given above, our coincidence rate~\textemdash{}
or modulus squared of the permanent~\textemdash{} is approximated
as the real part of a product of two \emph{independent sets of} samples,
$\mathbf{z},\tilde{\mathbf{z}}$, each composed of $E$ random vectors:
\begin{equation}
P_{N|M}\approx\Re\left[\left\langle p(\mathbf{M},\boldsymbol{z})\right\rangle _{E}\left\langle p(\mathbf{M},\tilde{\boldsymbol{z}})\right\rangle _{E}^{*}\right].\label{eq:PermS}
\end{equation}
The expression has two terms which are independent but conjugate on
average. The factored terms give independent estimates of the permanent
and its conjugate, and their product is an unbiased estimate of the
modulus squared. 

\subsection{Estimating the sampling error}

In any calculation using random sampling, it is essential to have
a statistical procedure for estimating the errors. A further modification
is therefore used in our numerical procedure to obtain error estimates.
Since we wish to understand how reliable the procedure is in terms
of its sampling error, the random sampling process described above
is repeated and averaged a large number of times for each sub-matrix,
to obtain independent statistical estimates. This allows the standard
deviation in the mean due to sampling error to be estimated. This
gives a theoretical error-bar in our graphs, to indicate how accurate
the calculation is. 

To achieve this in our calculations, we divide the total number of
ensembles $L$ into $L_{2}$ sub-ensembles, so that $L=L_{1}L_{2}$.
Each of the $L_{2}$ sub-ensembles has a large number $L_{1}$ of
independent noise terms. The final estimate is then an average over
the $L_{2}$ sub-ensemble estimates:
\begin{equation}
\left|\mathrm{perm}(\mathbf{M})\right|^{2}\approx\frac{1}{L_{2}}\Re\sum_{i=1}^{L_{2}}\left\langle p(\mathbf{M},\boldsymbol{z})\right\rangle _{L_{1}}^{(i)}\left\langle p(\mathbf{M},\tilde{\boldsymbol{z}})\right\rangle _{L_{1}}^{(i)*}.
\end{equation}
The notation $\left\langle \right\rangle _{L}^{(i)}$ indicates an
average over the $i$-th subensemble, as given in Eq.~(\ref{eq:subensemble-average}).
These individual averages have independent distributions. For large
$L_{1}$, by the central limit theorem, they are approximately Gaussian
since they involve sums of many independent random variables. In the
limit of small variances, the products of independent Gaussian distributed
quantities with a non-vanishing average remain nearly Gaussian distributed.
Hence each term of the sum over sub-ensembles is approximately Gaussian
distributed, as we are using independent estimates of the permanent
and its conjugate, or at worst has a $\chi^{2}$ distribution.

\begin{figure}
\begin{centering}
\includegraphics[width=0.9\columnwidth]{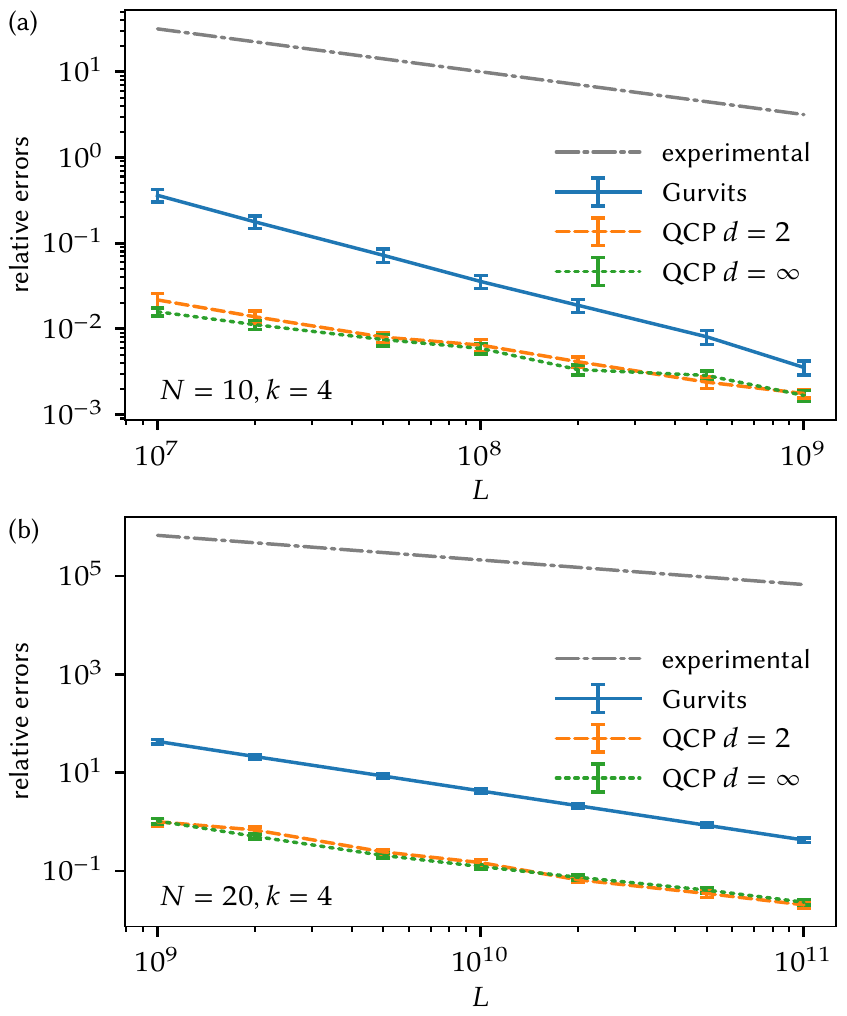}
\par\end{centering}
\caption{\textbf{\label{fig:relative-errors}}Actual error of the Gurvits method
(solid blue lines) and the QCP representation with $d=2$ (dashed
orange lines) and $d\rightarrow\infty$ (dotted green lines) for the
modulus-squared of the permanent, relative to the value of the permanent-squared,
as a function of the total number of samples $L=L_{1}L_{2}$. For
each point we have used $N_{\mathrm{m}}=100$ random unitary matrices
and $L_{2}=1000$ sub-ensembles of size (a) $L_{1}=10^{4}\dots10^{6}$
and (b) $L_{1}=10^{6}\dots10^{8}$. Here we consider $k=4$, (a) $N=10$
and (b) $N=20$. Dash-dotted grey lines denote the estimated experimental
error $\sqrt{\langle P\rangle_{\mathrm{e,m}}/L}$.}
\end{figure}

Combining sub-ensemble estimates in `quasi-conjugate' pairs that are
complex conjugate on average, but are independent, allows for an unbiased
estimate of the permanent squared. The use of sub-ensembles additionally
gives an approximate sampling error, using statistical methods valid
for Gaussian distributions. We have verified that these statistical
estimates are in fact reliable, by also calculating the error from
exact calculations of the permanent, where available. We find that
there is a good agreement between calculated and estimated errors,
and this agreement is maintained over a wide range of matrix sizes,
as shown in Fig.~\ref{fig:relative-errors}. Results are shown here
for $d=2$ and $d\rightarrow\infty$, with similar results for the
sampling errors, and good agreement with exact calculations. Since
the errors depend on the unitary, results are obtained here by averaging
over randomly chosen unitaries, as explained next.

\subsection{Unitary averages }

Although one can simulate individual unitaries with this approach,
it is more useful to know how the general performance scales with
network size. This is only meaningful when expressed as an average
over all possible unitary matrices. Serendipitously, infinite sums
over exponentially complex permanents are analytically calculable,
using results from the theory of matrix polar coordinates and random
matrices. This can be usefully employed to obtain the expected scaling
of the average count rate, for comparison purposes. We therefore can
express our results on computing permanents as averages over the Haar
measure of unitary matrices. These averaging techniques are well-known,
and can be applied to numerical averages.

In these calculations, we therefore obtain a set of randomly chosen
unitary matrices according to the Haar measure over the unitaries,
to give a \emph{third} level of random averaging. We repeat the above
process for each unitary matrix, to get unitary averages over the
sampling errors that occur in simulating typical random unitary matrices.

For comparison purposes, we note that a boson sampling experiment,
the probability of finding $N$ photons in the $M$ output modes of
a linear optical network is given by $P_{N|M}$. In the lossless case
where $T$ is unitary, this is the permanent squared of an $N\times N$
sub-matrix of an $M\times M$ unitary matrix, ${\rm perm}\left(\bm{U}_{N|M}\right)$~\cite{Tillmann2013}.
That is, in terms of our earlier notation, we take the lossless case
for simplicity ($t=1)$ and consider the transmission matrix of interest
as a unitary sub-matrix: $\bm{M}=\bm{U}_{N|M}$. We denote the average
over all unitary matrices of this permanent squared  as:
\begin{eqnarray}
\left\langle P_{N|M}\right\rangle _{\mathrm{U}} & = & \left\langle \left|{\rm perm}\left(\bm{U}_{N|M}\right)\right|^{2}\right\rangle _{\mathrm{U}}\nonumber \\
 & \equiv & \bar{P}_{N\vert M}.
\end{eqnarray}
 Using techniques from the theory of matrix polar coordinates and
unitary averages, this has a known scaling law given by~\cite{Arkhipov2012bBsonicBP,ScalingPerm}:
\begin{equation}
\ln\bar{P}_{N\vert M}=N\epsilon\left(k\right)+O(\ln N),\label{eq:ScalingLawBin}
\end{equation}
where $\epsilon\left(k\right)=k\ln k-(1+k)\ln(1+k)$ and $k=M/N$
is the channel ratio. 

Unitary averages that are carried out numerically involve a different
type of sampling error. Since the space of unitaries is extremely
high dimensional, this average is carried out in a Monte Carlo way,
by choosing random unitaries according to their Haar measure and averaging
over a finite set. In most cases we find that this variance over the
unitaries is rather small. Where it is significant, the unitary sampling
errors are plotted. This does not exclude the possibility that there
could be atypical unitaries where the count rates or estimation errors
are very different from the unitary average. However, these fall into
a set of very small measure, especially at large $N$ values.

\subsection{Sampling error comparisons}

There is a close relationship between the QCP method with qubits for
the permanent squared, and the Gurvits method for permanent approximation~\cite{gurvits2002deterministic},
in the case of an $N$-photon input and output. Both methods employ
random sampling over binary numbers. We extend this to qudit and continuous
cases, which has advantages in certain calculations. The Gurvits method
provides an unbiased estimate of the permanent. However, it introduces
systematic errors when computing the permanent squared, owing to effects
of the finite distribution width. It also has large sampling errors
when used for combinations of permanents, as described in the next
section. 

\begin{figure}
\begin{centering}
\includegraphics[width=0.9\columnwidth]{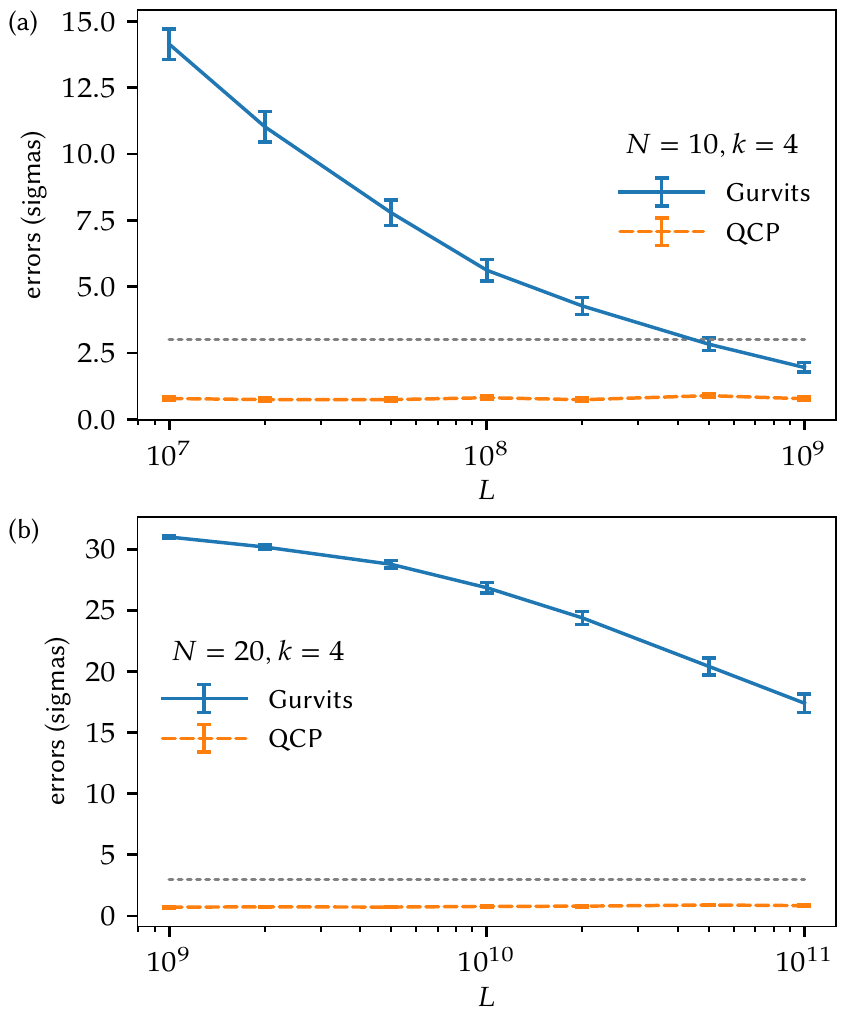}
\par\end{centering}
\caption{\textbf{\label{fig:errors-in-sigmas}}Actual error of the Gurvits
method (solid blue lines) and the QCP representation with $d\rightarrow\infty$
(dashed orange lines) for the modulus-squared of the permanent, expressed
in the units of the sampling error, as a function of the total number
of samples $L$. Graphs have $N=10$ and $N=20$ respectively, with
$k=4$. The dotted line indicates a $3\sigma$ confidence interval.
Other parameters are the same as in Fig.~\ref{fig:relative-errors}.}
\end{figure}

Figs.~\ref{fig:errors-in-sigmas}(a) and (b) show the systematic
errors of the modulus-squared of the permanent using the Gurvits method
and the QCP representation with $d\rightarrow\infty$ for a ratio
$k=4$, when averaged over $100$ randomly chosen $40\times40$ and
$80\times80$ unitary matrices respectively. These figures show that
the Gurvits algorithm has a large systematic bias for the permanent
squared. This is an order of magnitude larger than the standard deviation
in the examples given here. This statistical bias is reduced as the
number of samples is increased, but the samples required rapidly become
impractical as the matrix size $N$ is increased. 

The data processing for Fig.~\ref{fig:errors-in-sigmas} is performed
as follows. Each data point in the plot corresponds to $L$ total
samples, which is split in $L_{2}$ sub-ensembles of $L_{1}=L/L_{2}$
samples. For the Gurvits method the effective number of sub-ensembles
is $\tilde{L}_{2}^{(G)}=L_{2}\equiv10^{3}$, and for the QCP it is
$\tilde{L}_{2}^{(\mathrm{QCP})}=L_{2}/2$ (since two independent sub-ensembles
are used to produce a single sub-ensemble value of the permanent squared).
In this way we ensure that the total number of random samples used
for the comparison is the same for both methods. For each data point
we obtain a $N_{\mathrm{m}}\times\tilde{L}_{2}$ matrix $P_{i}^{(j)}$
of permanent squared values for each of $N_{\mathrm{m}}$ random matrices
and each sub-ensemble. We also calculate the exact permanent squared
values $\tilde{P}^{(j)}$ for each random matrix. 

Per-matrix actual errors are:
\begin{equation}
E^{(j)}=\left|\frac{1}{\tilde{L}_{2}}\sum_{i=1}^{\tilde{L}_{2}}P_{i}^{(j)}-\tilde{P}^{(j)}\right|\equiv\left|\langle P^{(j)}\rangle_{\mathrm{e}}-\tilde{P}^{(j)}\right|,
\end{equation}
 estimated sampling errors are calculated as:
\begin{equation}
E_{S}^{(j)}=\frac{\sqrt{\langle\left(P^{(j)}\right)^{2}\rangle_{\mathrm{e}}-\langle P^{(j)}\rangle_{\mathrm{e}}^{2}}}{\sqrt{L_{2}}}.
\end{equation}
We then plot the ratio $\Delta$ of actual errors $E$ relative to
the sampling errors $E_{\mathrm{samp}}$, that is, we calculate the
values
\begin{equation}
\Delta^{(j)}=E^{(j)}/E_{S}^{(j)},
\end{equation}
and plot the mean of the relative error $\langle\Delta\rangle_{\mathrm{m}}$
averaged over a finite set of unitaries and the estimated error of
the mean $\sqrt{\langle\Delta^{2}\rangle_{\mathrm{m}}-\langle\Delta\rangle_{\mathrm{m}}^{2}}/\sqrt{N_{\mathrm{m}}}$
as error-bars.

In summary, our methods gives an unbiased estimate of the permanent
squared, which is the relevant quantity in boson sampling experiments.
Even more importantly, our methods can also be used in the practical
case that the input is not a pure binary number state, or where the
output measurements are not $n$-th order correlations. 

From the mathematical viewpoint, these numerical methods unify the
complex P-representation approach in quantum optics with the computational
problem of efficient approximations of permanents and the permanent
squared, which is the relevant calculation for boson sampling.

\subsection{Experimental vs simulated errors}

To make useful \emph{predictions }about experimental observables,
perfect accuracy is not essential. It is only necessary to calculate
the output correlations with better than experimental errors. Such
errors can be accurately estimated from the count rates, which are
determined by the permanents in the idealized case of number-state
inputs. In the experimental case, the measured standard deviations
are simply Poissonian errors in the counts, whose scaling is estimated
in the figures. 

To make sensible comparisons, one must choose time-scales that are
comparable. Here we note that both digital computers and photonic
devices use similar counting and logic electronics, which limit the
speed of any one simulation or measurement task. To obtain good statistics,
both types of device require repeated measurement and averages over
many samples, to reduce sampling errors.

The time taken is then just $T=LT_{N}$ for $L$ samples, where $T_{N}$
is the time taken per sample for $N$ channels. We assume that the
factor $T_{N}$ is similar for digital and photonic devices. In both
cases, there are other effects due to computational overheads and/or
physical limits as $N$ increases, which are neglected here for simplicity.
Given this assumption, we can compare the sampling errors over similar
time-scales, that is, with the same number of samples. Our main purpose
is to show that, at a given error, the computation of quantum correlations
is as feasible as an experimental measurement.

\begin{figure*}
\begin{centering}
\includegraphics{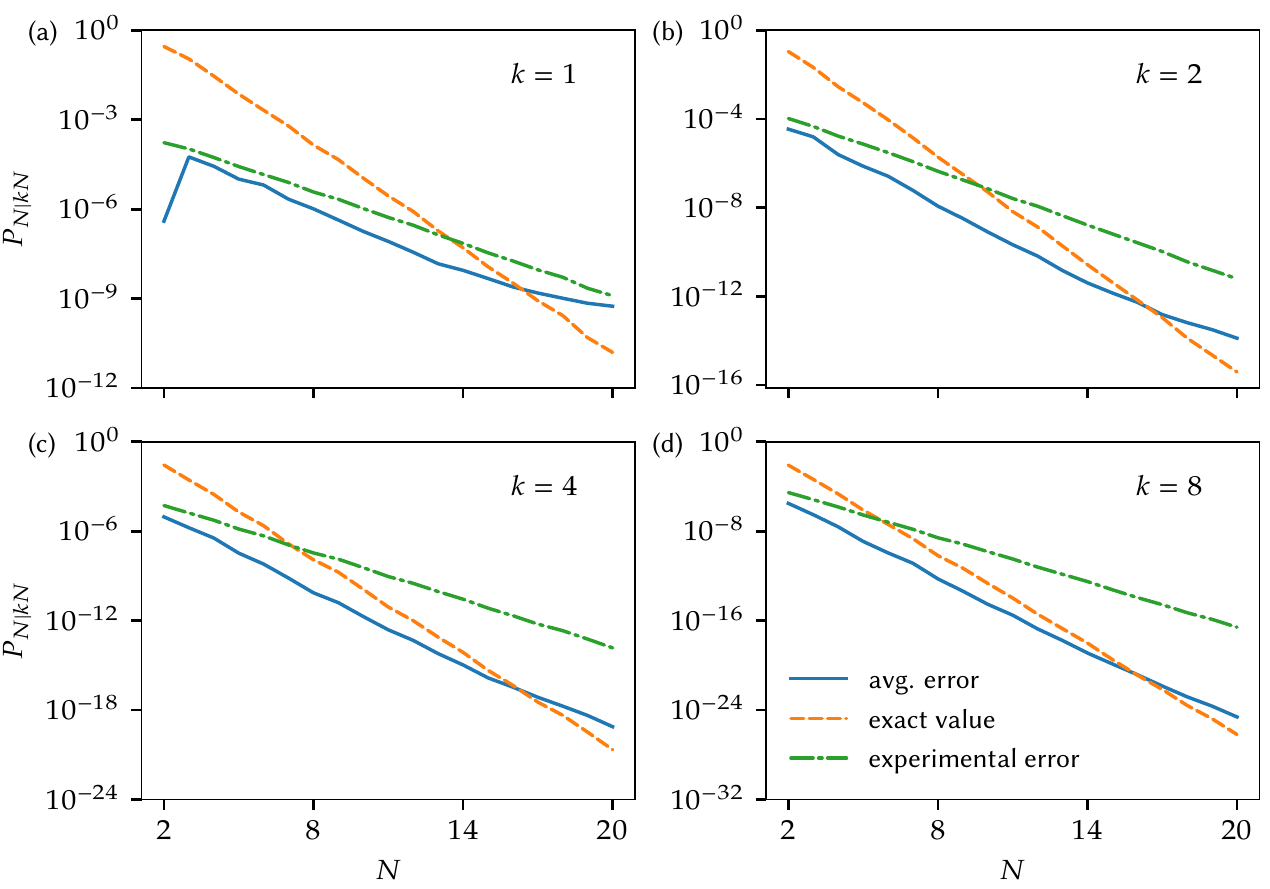}
\par\end{centering}
\caption{\label{fig:Permanents_Scaling}Estimation of the error for unitary-averaged
coincidence rate  $\bar{P}_{N\vert kN}$ as a function of $N$ using
the QCP method with a random discrete phase $d\rightarrow\infty$.
For each point we have used $N_{\mathrm{m}}=100$ random unitary matrices
and $L_{2}=100$ ensembles of $L_{1}=10^{5}$ samples each. Dashed
orange line corresponds to the average for the exact value of $P_{N\vert kN}$.
Solid blue line is the average error $\epsilon$, compared to the
exact value.    Dash-dotted green line denotes the estimated experimental
error $\sqrt{\langle P\rangle_{\mathrm{e,m}}/L}$. \textbf{This is
always greater than the simulation error, for the same number of samples.}
Here we consider: (a) $k=1$, (b) $k=2$, (c) $k=6$ and (d) $k=10$.
}
\end{figure*}

Figs.~\ref{fig:Permanents_Scaling}(a)-(d) show the average modulus
squared of the permanent of an $N\times N$ sub-matrix, denoted by
$\bar{P}_{N\vert kN}$, as well as the mean error, or average deviation
of the sampled result from the exact value. The mean error $E$ is
defined as $E=\langle|P-\langle\tilde{P}\rangle_{\mathrm{e}}|\rangle_{\mathrm{m}}$,
where $P$ is exact, and $\langle\tilde{P}\rangle_{\mathrm{e}}$ is
the quantum simulation  ensemble average. All results are averaged
over a finite unitary ensemble of matrices $\langle\rangle_{\mathrm{m}}$.
We have also plotted the Poissonian experimental error, which is asymptotically
\emph{larger} than the error of our simulations for an identical sample
number. 

While exact results for permanents are not available at large \emph{N,
}owing to their complexity\emph{,} the average\emph{ }scaling behavior
of $N$-channel coincidences over all possible unitaries can be calculated
analytically. One could make comparisons of every possible unitary,
but as they are uncountably infinite in number, this approach is not
very practical. In each experiment with $E$ trials, the unitary average
Poissonian measurement variance due to shot-noise is $\left\langle \sigma^{2}\right\rangle _{\mathrm{U}}=\bar{P}/L$,
which therefore scales as $\ln\left\langle \sigma^{2}\right\rangle _{\mathrm{U}}=N\epsilon\left(k\right)-\ln L+O(\ln N)$.

To verify an experimental probability, one must have a theoretical
error less than this experimental sampling error. Typically the channel
ratio $k$ ranges from unity~\textemdash{} its minimum value~\textemdash{}
up to a number of order $N$, which has theoretical advantages in
computability theory~\cite{Aaronson2011}. Hence our numerical results
given below are compared with sampling errors in the corresponding
experiments, over a range of parameters. The results show that the
computational error using our method has a \emph{better} scaling than
experiment, for the same size $L$ of the ensemble of measurements
or samples. 

The graphical comparisons demonstrate a necessary requirement for
any assessment method. Without this favorable scaling, the calculation
of expected correlations would take exponentially longer than the
experiment itself. Instead, computational sampling errors reduce
rapidly with $N$, faster than experimental error reductions. This
means that the computational time is not prohibitive.

However, the graphs also show that the average permanent values reduce
even faster. This makes direct permanent\emph{ measurements} problematic
at large $N$, except for matrices with large permanents. In other
words, at sufficiently large $N$ the average count rate for any individual
set of channels goes rapidly to zero for most unitary sub-matrices,
so that the measurement of an individual correlation or moment is
impractical. We turn to a solution of this problem in the next section.

\section{Correlation based verification strategy\label{sec:combined-correlations}}

As boson sampling experiments improve, the problem of verification
at large $N$ will become acute~\cite{Walschaers2016_NJP18,Spagnolo2014,Carolan2014,Aaronson:2014,Tichy2014,Bentivegna2015,AolitaEisert2015}.
One needs a verification protocol that has the following properties.
It should be:
\begin{enumerate}
\item Calculable at large $N$, in practical timescales.
\item Measurable at large $N$, in practical timescales.
\item Able to distinguish the required distribution from other ones.
\item Applicable to all possible transmission unitaries. 
\item Unable to be readily mimicked.
\item Part of a well-defined progression of tests.
\end{enumerate}
The last condition is based on the assumption that no single test
will be conclusive. Since the exact permanents are exponentially hard
to compute, a test that relies on knowing the exact permanent for
any sub-unitary will not satisfy the first criterion. Verification
methods that rely on calculating single permanents exactly have a
limited applicability. They may work well at small $N$ values. Eventually,
exponential complexity will make such methods impractical. Therefore,
other methods must be found.

Even our quantum simulation techniques cannot reliably predict the
permanent squared with high enough accuracy for all unitaries. At
large $N$, it is really only useful for permanents that have large
values. However, it is the low experimental count rate that is the
main limiting factor. The methods described here can compute any permanent
that is measurable with \emph{better} than the experimental error.
An ideal solution is to have an assessment signature that is both
measurable and calculable at large $N$ values. We turn to this challenging
question in this section.

\subsection{Combined correlations}

\begin{figure}
\includegraphics{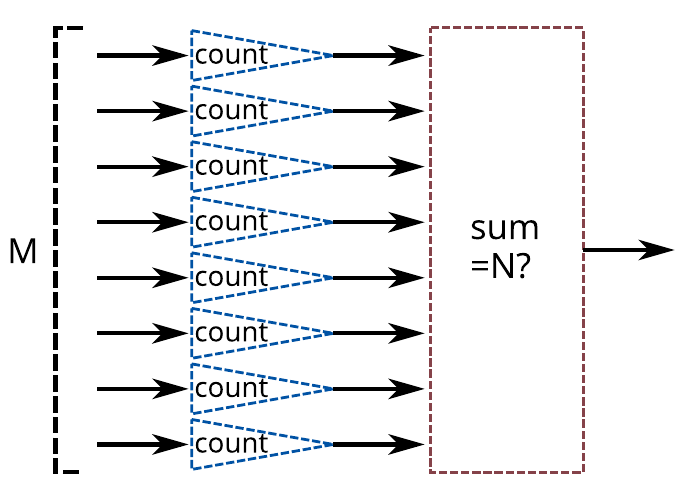}

\caption{\label{fig:combined-correlations-scheme}A schematic of the combined
correlations strategy. Photo-detectors on output channels are indicated
by triangles. Counter outputs must be binary or passed through a step
function filter. Thus, channel counts are always $0$ or $1$, even
for multiphoton events, which give total counts less than $N$. As
a result, the final output counts all events with exactly $N$ photons
in $N$ distinct channels. This sums over all possible channel combinations.}
\end{figure}

To obtain a quantity that is both measurable and calculable for typical
unitaries at large $N$, we use the fact that collecting data from
a single combination of $N$ output channels is extremely inefficient.
Almost all the output information is lost with this procedure. Yet
both the experimental data and the simulation data can provide parallel
information about all channel combinations simultaneously, utilizing
far more of the available information.

Hence, we first consider the \emph{combined} correlation $C_{N|M}\equiv\langle\hat{C}_{N|M}\rangle$,
defined as the sum over all different channel combinations $\sigma$
of length $N$:
\begin{equation}
\hat{C}_{N|M}=\sum_{\sigma}\prod_{j\in\sigma}^{N}\hat{n}_{j}.\label{eq:Correlations}
\end{equation}

These sums combine an exponentially large number of permanents, each
of which is exponentially hard to compute. Despite this, they can
be evaluated efficiently with the QCP method using a modification
of this technique that employs a discrete Fourier transform (DFT);
see~the Appendix for details. While Gurvits type binary methods could
also be used in combination with the DFT approach, the sampling error
is much greater than with a qudit or continuous sampling, in addition
to the systematic error problem already identified with such methods. 

This combined channel, continuous sampling method provides an exceptionally
efficient route to the randomized calculation of \emph{all} the exponentially
large number of $N$-th order correlations, each involving an exponentially
large permanent calculation. The advantage of using $C_{N|M}$ as
the assessment signature is that it has high count rates even at large
$N$ values, and therefore is scalable. This can be measured experimentally
with high average count rates, especially in the important large $k$
regime~\cite{Arkhipov2012bBsonicBP,ScalingPerm}. A possible experimental
realization would be to attach a detector to every output of the Boson
Sampling device, triggered if it detects any photons. The outputs
of these detectors are joined in a correlator which registers an event
when exactly $N$ of the detectors are triggered. See Fig.~\ref{fig:combined-correlations-scheme}
for an illustration of the strategy.

Naturally, this is still not exact for finite resources, due to sampling
errors both in the experiment measurements and theoretical estimates.
This leads to a problem: for large $N$, $C_{N|M}$ for any unitary
is almost always given by its unitary average. This is known as the
self-averaging property of a large random unitary. 

Although satisfying the other criteria, \foreignlanguage{english}{$C_{N|M}$}
therefore doesn't adequately distinguish between the different unitaries,
due to sampling errors in the measurement process, combined with self-averaging.
A fraudulent boson sampling device could be constructed to approximately
replicate the required statistics, without having to process any information
about the unitary. 

\subsection{Channel-deletion verification\label{sec:channel-deleted}}

\begin{figure}
\includegraphics{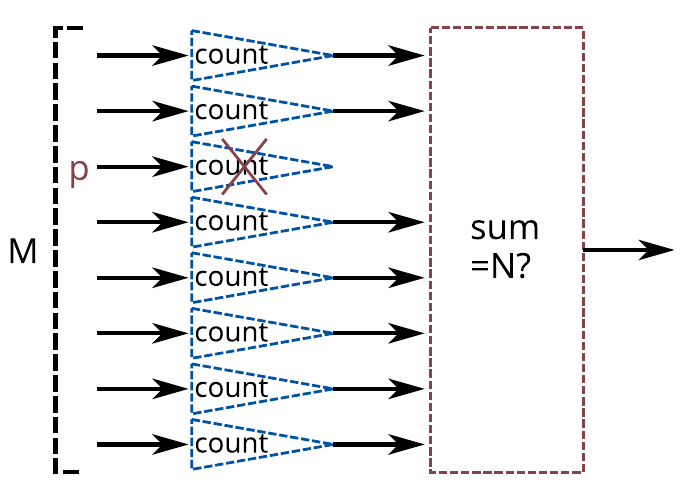}

\caption{\label{fig:combined-correlations-excluded-scheme}A schematic of the
channel deleted combined correlations strategy. One or more channels
are deleted randomly by switching off their counters. Therefore, only
the events with\emph{ N }counts in $N$ distinct channels, excluding
deleted channels, are counted. This gives a high total count rate.
At the same time, it provides a successively more detailed fingerprint
of the unitary. This becomes effectively unique as the number of deleted
channels is increased. }
\end{figure}

An improved strategy is needed in order to discriminate between different
unitary matrices, and obtain a unique signature of the required permanent
distribution. Therefore, to assess the boson sampling device, we now
consider a hierarchy of experiments in which one or more channels
is deleted from the channel combinations. This leads to channel-deleted,
combined correlation $C_{N|M}^{(p)}$, which sum over all $N$-fold
correlations that \emph{don't} include a specific channel $p$, as
illustrated in Fig.~\ref{fig:combined-correlations-excluded-scheme}:
\begin{equation}
\hat{C}_{N|M}^{(p)}=\sum_{\sigma,p\notin\sigma}\prod_{j\in\sigma}^{N}\hat{n}_{j}.\label{eq:CorrDC}
\end{equation}

In this approach we measure the combined permanents conditioned on
channel $p$ having no counts. Similarly, one can have a hierarchy
of measurements $C_{N|M}^{(\rho)}$, which are conditioned on channels
in the set $\rho=\left\{ p_{1},p_{2},\ldots\right\} $ having no counts.
Eventually this exhaustively enumerates all measurable $N$-th order
correlations, starting with the most readily measured combinations
having the highest count rates and the lowest experimental errors. 

The advantage of this approach is that the goal of a boson sampling
device is to generate samples with a permanent distribution. However,
any probability distribution over a finite range has a unique fingerprint~\cite{Moran}:
the set of all its moments. The hierarchy of channel-deleted combinations
converges to this unique signature, in the limit in which all possible
deletions are included.

\begin{figure}[h]
\begin{centering}
\includegraphics{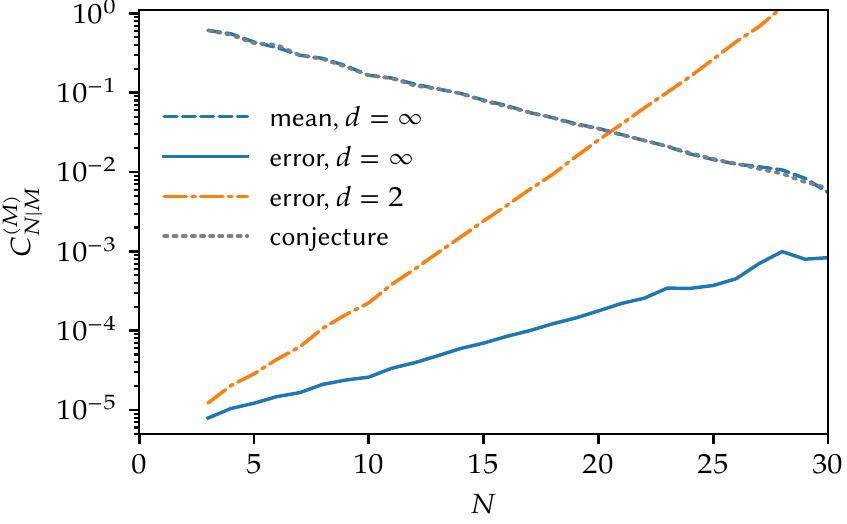}
\par\end{centering}
\caption{\label{fig:Correlations-mean}Combined correlations $C_{N|M}^{(M)}$
given in Eq.~(\ref{eq:CorrelationQCP}) evaluated using the QCP method.
Here we have used $d\rightarrow\infty$, $k=6$, $N_{\mathrm{m}}=1$
and $L_{2}=10^{4}$ ensembles of $L_{1}=10^{6}$ samples. The dotted
grey line is the analytical result given in Eq.~(\ref{eq:ConjectureC}),
where $N=M/k$. The dashed blue line corresponds to our numerical
results. The solid blue line is the average estimate of the error
in the mean $E_{S}=\langle\sqrt{\langle C^{2}\rangle-\langle C\rangle^{2}}\rangle_{\mathrm{m}}/\sqrt{L_{2}}$.
The dash-dotted orange line is the estimate of the error for the case
of $d=2$.}
\end{figure}

\begin{figure}[h]
\begin{centering}
\includegraphics{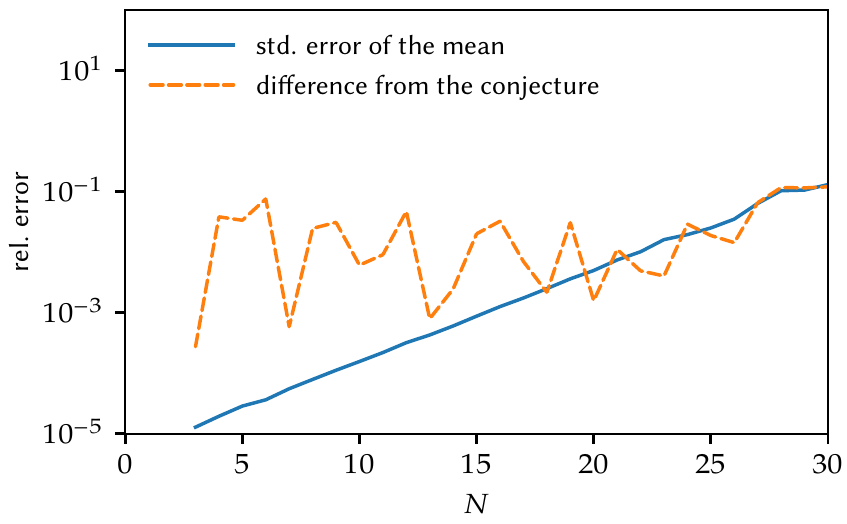}
\par\end{centering}
\caption{\label{fig:Correlations-error}Modulus of the difference between the
numerically obtained combined correlation, and the conjecture $C_{N|M}^{(M),\mathrm{conj}}$
(dashed orange line) plotted against the error estimate from Fig.~\ref{fig:Correlations-mean},
normalized on $C_{N|M}^{(M),\mathrm{conj}}$ (solid blue line). All
other parameters as in Fig.~\ref{fig:Correlations-mean}.}
\end{figure}

\begin{figure}[h]
\begin{centering}
\includegraphics{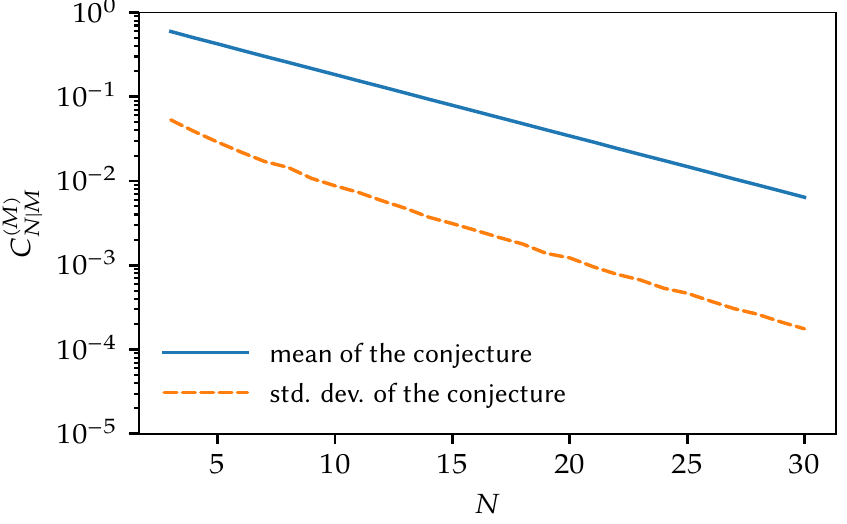}
\par\end{centering}
\caption{\label{fig:Correlations-conjecture}Standard deviation of the analytic
conjecture $C_{N|M}^{(M)\mathrm{,conj}}$ for $10^{4}$ random matrices
(dashed orange line) plotted against its mean value (solid blue line),
showing that there is a substantial variation between unitaries even
for relatively large $N$. All other parameters as in Fig.~\ref{fig:Correlations-mean}.}
\end{figure}

Calculation with complex-P distributions is straightforward. The specified
channel becomes a loss reservoir for the remaining $M-1$ channels.
This modified correlation can be readily evaluated using the QCP method,
since its expectation value can be calculated probabilistically as:
\begin{equation}
C_{N|M}^{(p)}=\left\langle \frac{1}{M-1}\sum_{j}e^{-\mathrm{i}jk\delta}\prod_{k=1}^{M-1}\left[1+\mathrm{e}^{\mathrm{i}j\delta}m_{k}\right]\right\rangle .\label{eq:CorrelationQCP}
\end{equation}
Here $m_{k}$ is a scaled boson number. We note, however, that even
this technique has too large a sampling error for large enough $N$
values. Eventually, it will become essential to obtain an analytically
calculable signature to verify boson sampling for the large $N$ case,
which is important from the complexity theory viewpoint.

For this modified correlation with a single channel-deletion we give
a physical argument that allows to conjecture an analytical form at
large $N$. In this case the selected channel $p$ acts as a loss
reservoir for the other $M-1$ channels. Therefore, we can write $\alpha^{\mathrm{(out)}}=U\alpha^{\mathrm{(in)}}$,
$\beta^{\mathrm{(out)}}=\beta^{\mathrm{(in)}}U^{\dagger}$ so that
\begin{eqnarray*}
\left\langle n_{j}^{\mathrm{(out)}}\right\rangle  & = & \left\langle U\alpha^{\mathrm{(in)}}\beta^{\mathrm{(in)}}U^{\dagger}\right\rangle _{jj}=\sum_{i=1}^{N}\left|U_{ij}\right|^{2}=T_{j},
\end{eqnarray*}
 and $\sum_{j=1}^{M}T_{j}=N$. Next, we exclude the counts in channel
$p$, since this acts as a loss reservoir. From unitarity, $\sum_{j=1}^{M}T_{j}=Nt$
, and we get:
\[
\sum_{j\neq p}^{M}T_{j}=Nt-T_{p}=Nt\left(1-T_{p}/Nt\right)=t_{p}N,
\]
where the effective channel loss rate for channel $p$ is $t_{p}=t\left(1-\sum_{i=1}^{N}\left|U_{pi}\right|^{2}/N\right)$
.

For large $k\equiv M/N$, the modified correlation~\textemdash{}
which is the probability of $N$ coincidences in any of the remaining
$M-p$ channels~\textemdash{} is given by the sum of permanents of
the $N\times N$ sub-matrices of the remaining $\left(M-1\right)\times\left(M-1\right)$
matrix. We conjecture that this is given asymptotically by taking
an $M-1$ dimensional unitary average, together with an additional
reduction of $t_{p}^{N}$ due to channel loss, in channel $p$, so
that:
\begin{equation}
C_{N|kN}^{(p),\mathrm{conj}}\,\mathop{\sim}\limits _{k\to\infty}\,\,\frac{t_{p}^{N}(kN-1)!(kN-2)!}{((k-1)N-1)!((k+1)N-2)!},\label{eq:ConjectureC}
\end{equation}
 We have tested this result using our simulation methods, that allow
us to sum over exponentially large numbers~\textemdash{} up to $10^{34}$~\textemdash{}
of large permanents in parallel. In Figs.~\ref{fig:Correlations-mean}
and~\ref{fig:Correlations-error} we show numerical results for the
channel combinations $C_{N|M}^{(p)}$ using the QCP method as well
as for the conjecture $C_{N|M}^{(p),\mathrm{conj}}$. This remarkable
task of summing over exponentially large numbers of quantities, each
of which is exponentially difficult to compute, demonstrates the versatility
of the simulation techniques used here. 

Having established that the conjecture gives asymptotically correct
results~\textemdash{} with the exception of a set of unitaries of
measure close to zero, such as unitaries near to the identity~\textemdash{}
one might inquire whether it is able to distinguish between the different
random unitaries. This is investigated in Fig.~\ref{fig:Correlations-conjecture},
which plots the mean and standard deviation of the calculated last-channel-deleted
count rate, $C_{N|M}^{(M),\mathrm{conj}}\,$, for $28$ randomly chosen
sets of $10^{4}$ unitaries with $k=6$. The largest unitaries considered
are $180\times180$. This graph shows that, while the standard deviation
of the count-rate due to the randomness of the unitaries is less than
the rate itself, the deviation from one unitary to the next is far
from being negligible. 

Our numerical simulations show that the conjecture is asymptotically
valid at large $k$ for a single unitary, apart from anomalous matrices
with zero asymptotic measure, such as the unit matrix. It is in excellent
agreement with our quantum simulations of a random unitary, to a relative
accuracy of order $1\%$ for $k=6$. This signature is purely analytic,
and calculable for large sizes. It is experimentally accessible, and
can distinguish unitaries. It can be generalized recursively to allow
increasing numbers of channel deletions, giving an increasingly unique
fingerprint of the relevant permanent squared distribution, so that
it will eventually distinguish any unitaries. Of course, such distinction
is only possible if the unitaries are sufficiently distant relative
to the sampling error.

There have been many different approaches to assess boson sampling
experiments. We must start by ruling out techniques that require an
exact calculation of the permanent for an arbitrary matrix. This is
a \#P hard computation and hence cannot be carried out for large $N$
with known algorithms. Measurement of the permanent within multiplicative
error is similarly hard (on average), due to exponential suppression
of count rates. 

There are interesting methods that distinguish the boson sampling
network from an uniform sampler~\cite{Spagnolo2014,Carolan2014}.
Other tests are based on considering either distinguishable or indistinguishable
photons on the boson sampler~\cite{Spagnolo2014,Carolan2014,Aaronson:2014}
or distinguishing particles based on their statistics~\cite{Walschaers2016_NJP18,Shchesnovich:2016_PRL}.
In~\cite{Shchesnovich:2016_PRL} an assessment protocol using the
generalized bunching effect for bosons was proposed. In this case,
for indistinguishable photons an absolute maximum of the probability
of detecting all input particles in some output modes exists. This
assessment protocol has the advantage that is detectable in a polynomial
number of runs for an arbitrary network, but it is not clearly able
to distinguish valid from all other possible invalid output distributions.
These are useful tests, but generally will not distinguish between
all possible distributions, and so do not satisfy criterion 3.

Many-particle interference can also be used as an assessment for boson
sampling~\cite{Walschaers2016_NJP18} using a mode correlator, which
contains the statistical properties of the particles. Using Random
Matrix theory it is possible to obtain analytical result for the first
three moments for certain conditions on the sub-matrix that contains
the information of the coupling between the input and output modes.
We note that such low order moments are not used in our assessment
protocol, which instead relies on high-order moments that are directly
relevant to the boson sampling problem. Recent experiments that consider
if the input photons are distinguishable or indistinguishable photons
have been performed based on machine learning~\cite{Agresti:2017}
or Sylvester matrices~\cite{Viggianiello:2017}. Generally speaking,
these methods also do not seem to distinguish different unitaries,
although this is a nontrivial requirement.

In~\cite{Tichy2014} a method based on symmetric sampling matrices
was proposed. This distinguishes a boson sampler from another device
as a mean field sampler. The test relies on verifying the suppression
law for Fourier matrices~\cite{Tichy2010}. It has the advantage
that the proposal can be carried out for any $N$, which makes it
extremely powerful. There is no doubt that this is a very sensitive
test. Yet, as it does not allow assessment for any other unitary,
it cannot satisfy criterion 4. 

Our proposed method has the advantage that it can, in principle, treat
any type of matrices, not just Fourier-transform unitary ones. It
also does not rely on the zero-transmission law that uses the Fourier
transform. Our proposal treats $N-$th order correlations, but does
not rely on a generalized average bunching effect. 

Finally, we note that counting rates are rather crucial at large $N$
values. Efficient nanowire detectors with $t=0.9$ (using an optimistic
estimate) and $1\,\mathrm{GHz}$ measurement rates~\cite{Takesue2015}
could allow one to reach high count rates. Coincidence rates as high
as one per second even for $N=100$ and $k=10$, are not impossible,
making these tests possible for large networks. Even very large boson
sampling experiments beyond the limit of exact computability for permanents
are potentially viable for our test. Such large $N$values would give
an example of a quantum computation of a random bitstream that is
inaccessible with digital computers. It is important to note that
our complex phase-space methods do \textbf{not} replicate the random
bitstream of coincidence counts, so they cannot replace the quantum
network. Their role is to enable the design and assessment of the
device. 

\section{Conclusion\label{sec:Conclusion}}

In summary, we obtain novel quantum simulation methods based on complex
phase-space P-distributions. These can be used for the efficient simulation
of completely general linear photonic networks, using samples of integrals
over coherent states. Our method is applicable to arbitrary inputs,
losses and outputs. The generation of each sample is possible within
polynomial time in $N$ and $M$. By contrast, classical generation
of random photon number counts takes an exponentially large time in
$N$ for each sample, as we expect from known fundamental complexity
properties of boson sampling. Thus, our sampling technique is exponentially
faster than generating photon counts.

We also calculate how the sampling error in an estimated counting
probability scales with the number of samples. Here there is an unexpected
result. Rather than giving an exponentially slower method than experiment,
as one would obtain using classical random bit generation, the algorithm
is exponentially \emph{faster} than experiment. This is because of
the properties of the coherent basis. Weighted contour integral sampling
of moments has a much lower error at large $N$ values than estimating
moments from numbers of photon counts. This makes possible the prediction
of any photonic correlation, with less error than an experimental
measurement taking a similar time.

In a very large network~\textemdash{} the target scenario for boson
sampling~\textemdash{} the limitation of any assessment protocol
is the low experimental count rate in any given set of channels. This
is especially an issue when one is no longer able to efficiently compute
permanents using standard methods. Accordingly, we propose a combined
channel grouping protocol that allows one to assess large-scale boson
sampling experiments. This uses a channel deletion protocol to distinguish
different unitaries with relatively low experimental sampling error.
The number of deleted channels can be recursively increased for more
and more fine-grained distinction between the possible distributions.
In such cases, one could then claim validation at successively more
challenging levels of channel deletion.

We emphasize that our quantum simulation results are still limited
by sampling errors. This is a strong limitation, even with the relatively
efficient methods we use. However, these sampling errors are unbiased,
and are generally much less than with other techniques carried out
with the same computational time requirement. In addition, since the
analytic test we propose is a conjecture, its limitations need to
be investigated by further studies using random matrix theory. Application
of the simulations to dissipation and noise in boson sampling interferometry
will be given elsewhere. Finally, an interesting open question is
the applicability of these methods to other input states.

\section*{Acknowledgments}

This research was supported in part by the National Science Foundation
under Grant No. NSF PHY-1125915, and by the Australian Research Council. 

\appendix

\section{Combined correlation calculations\label{sec:Combined-correlations}}

We wish to use the QCP complex phase-space method in order to evaluate
the sums of combined correlations $C_{N|M}\equiv\langle\hat{C}_{N|M}\rangle$
shown in Fig.~\ref{fig:combined-correlations-scheme}, given by:
\[
\hat{C}_{N|M}=\sum_{\sigma\in S_{N}}\prod_{j\in\sigma}\hat{n}_{j},
\]
where $S_{N}$ is the set of combinations of length $N$ of integers
from $1$ to $M$ ($\left\{ 1,2,\dots,N-1,N\right\} $, $\left\{ 1,2,\dots,N-1,N+1\right\} $,
$\dots$, $\left\{ M-N+1,M-N+2,\dots M-1,M\right\} $). This set corresponds
to all different $N$-th order correlations of the output channels.
The expectation value of this operator combines an exponentially large
number of permanents of $N\times N$ sub-matrices of an $M\times M$
unitary matrix, \emph{each} of which is exponentially hard to compute. 

We first consider the following correlation polynomial, defined for
$j=0,\ldots M-1$:
\[
\hat{D}_{j}=\prod_{k=1}^{M}\left[1+\mathrm{e}^{\mathrm{i}j\delta}\hat{n}_{k}\right],
\]
with $\delta\equiv2\pi/M$. There are $M$ distinct correlation polynomials
for $j=0,\ldots M-1$, each including all possible combinations of
$\hat{n}_{k}$. Their Fourier transform is given by:
\[
\hat{F}_{k}=\frac{1}{M}\sum_{j=1}^{M}\mathrm{e}^{-\mathrm{i}jk\delta}\hat{D}_{j}.
\]
Here $\hat{D}_{j}$ includes all possible multinomials in $\hat{n}_{k}$.
For $N-$th order multinomial terms, the phase factors cancel, giving
us the required correlation:
\[
\hat{F}_{N}\equiv\hat{C}_{N|M}.
\]

Using the results of Section~\ref{sec:Discrete-complex-P}:

\begin{eqnarray*}
\langle\hat{C}_{N|M}\rangle & = & \frac{1}{M}\sum_{j=1}^{M}\mathrm{e}^{-\mathrm{i}jN\delta}\langle\hat{D}_{j}\rangle\\
 & = & \frac{1}{M}\sum_{j=1}^{M}\mathrm{e}^{-\mathrm{i}jN\delta}\langle\prod_{k=1}^{M}\left[1+\mathrm{e}^{\mathrm{i}j\delta}\hat{n}_{k}\right]\rangle\\
 & = & \frac{1}{M}\sum_{j=1}^{M}\mathrm{e}^{-\mathrm{i}jN\delta}\langle\prod_{k=1}^{M}\left[1+\mathrm{e}^{\mathrm{i}j\delta}n_{k}^{(o)}\left(\bm{q},\tilde{\bm{q}}\right)\right]\rangle_{P}
\end{eqnarray*}
where $n_{k}^{(out)}$ is given by Eq~(\ref{eq:photon-number-variable})
as:

\[
n_{k}^{(out)}=r^{2}\left(\sum_{s\in\sigma}T_{ks}z^{q_{s}}\right)\left(\sum_{t\in\sigma}T_{kt}z^{\tilde{q}_{t}}\right)^{*}
\]
for an $M\times M$ transmission matrix $\bm{T}$, and single-photon
inputs in a set of distinct channels $\sigma\in S_{N}$. Here $z=e^{2\pi i/d}$
and $q_{s}$ and $\tilde{q}_{t}$ are random integers uniformly distributed
in the range $\left[0,d-1\right]$. The $P$-function used in the
averaging $\langle\rangle_{P}$ is the product of the factors~(\ref{eq:single-mode-P-function})
for one-photon inputs and ones for zero-photon inputs:

\[
P\left(\bm{q},\tilde{\bm{q}}\right)=\prod_{m=1}^{N}\frac{1}{r^{2N}}z^{\tilde{q}_{m}}z^{-q_{m}}
\]

Denoting $\alpha_{k}=\sum_{s\in\sigma}T_{ks}z^{q_{s}}$, $\beta_{k}^{*}=\sum_{j\in\sigma}T_{kt}z^{\tilde{q}_{t}}$,
and introducing a scaled boson number $m_{k}=\alpha_{k}\beta_{k}$,
and a combined unit random $y_{m}=z^{\tilde{q}_{m}}z^{-q_{m}}$, we
get for the combined correlation:

\begin{eqnarray*}
\langle\hat{C}_{N|M}\rangle & = & \frac{1}{M}\sum_{j=1}^{M}\mathrm{e}^{-\mathrm{i}jN\delta}\times\\
 &  & \quad\langle\left(\prod_{m=1}^{N}\frac{1}{r^{2N}}y_{m}\right)\left(\prod_{k=1}^{M}\left[1+\mathrm{e}^{\mathrm{i}j\delta}r^{2}m_{k}\right]\right)\rangle.
\end{eqnarray*}
Since all the terms with more or less than $N$ of $m_{k}$ factors
are eliminated by the Fourier transform technique, the only ones that
are left have the $r^{2N}$ multiplier that annihilates the $1/r^{2N}$
of the P-function:

\[
\langle\hat{C}_{N|M}\rangle=\frac{1}{M}\sum_{j=1}^{M}\mathrm{e}^{-\mathrm{i}jN\delta}\langle\left(\prod_{m=1}^{N}y_{m}\right)\left(\prod_{k=1}^{M}\left[1+\mathrm{e}^{\mathrm{i}j\delta}m_{k}\right]\right)\rangle.
\]

\section{Combined correlations with excluded channels}

We now consider how to distinguish unitaries. Define a count rate
$C_{N|M}^{\left(\rho\right)}$, such that we definitely have no count
in the set $\rho$ consisting of $Q$ different channels, and we don't
care where the other counts are except that there are $N$ single
counts in total. Define $S_{N}^{\left(\rho\right)}$ as the set of
all $N$ channel combinations that exclude the channels from the set
$\rho$. 

Similarly to the previous section, we have

\[
\hat{C}_{N|M}^{\left(\rho\right)}=\sum_{\sigma\in S_{N}^{\left(\rho\right)}}\prod_{j\in\sigma}\hat{n}_{j}.
\]
Consider the modified correlation polynomial, defined for $j=0,\ldots M-Q-1$:
\[
\hat{D}_{j}^{(\rho)}=\prod_{k=1\dots M,\ k\notin\rho}\left[1+\mathrm{e}^{\mathrm{i}j\delta_{Q}}\hat{n}_{k}\right],
\]
with $\delta_{Q}\equiv2\pi/\left(M-Q\right)$. The Fourier transform
is given by:
\[
\hat{F}_{k}^{(\rho)}=\frac{1}{M-Q}\sum_{j=1}^{M-Q}\mathrm{e}^{-\mathrm{i}jk\delta_{Q}}\hat{D}_{j}^{(\rho)}.
\]
And as before, for $N-$th order multinomial terms, the phase factors
cancel, giving us the required correlation:
\[
\hat{F}_{N}^{(\rho)}\equiv\hat{C}_{N|M}^{(\rho)}.
\]

Similarly to the previous section, the probabilistic formula for the
expectation value of $\hat{C}_{N|M}^{(\rho)}$ can be found to be

\begin{eqnarray*}
\langle\hat{C}_{N|M}^{(\rho)}\rangle & = & \frac{1}{M-Q}\sum_{j=1}^{M-Q}\mathrm{e}^{-\mathrm{i}jN\delta_{Q}}\times\\
 &  & \quad\langle\left(\prod_{m=1}^{N}y_{m}\right)\left(\prod_{k=1\dots M,\ k\notin\rho}\left[1+\mathrm{e}^{\mathrm{i}j\delta_{Q}}m_{k}\right]\right)\rangle.
\end{eqnarray*}

\subsection*{}

\bibliographystyle{apsrev4-1}
\bibliography{BosonSampling}

\begin{thebibliography}{58}%
\makeatletter
\providecommand \@ifxundefined [1]{%
 \@ifx{#1\undefined}
}%
\providecommand \@ifnum [1]{%
 \ifnum #1\expandafter \@firstoftwo
 \else \expandafter \@secondoftwo
 \fi
}%
\providecommand \@ifx [1]{%
 \ifx #1\expandafter \@firstoftwo
 \else \expandafter \@secondoftwo
 \fi
}%
\providecommand \natexlab [1]{#1}%
\providecommand \enquote  [1]{``#1''}%
\providecommand \bibnamefont  [1]{#1}%
\providecommand \bibfnamefont [1]{#1}%
\providecommand \citenamefont [1]{#1}%
\providecommand \href@noop [0]{\@secondoftwo}%
\providecommand \href [0]{\begingroup \@sanitize@url \@href}%
\providecommand \@href[1]{\@@startlink{#1}\@@href}%
\providecommand \@@href[1]{\endgroup#1\@@endlink}%
\providecommand \@sanitize@url [0]{\catcode `\\12\catcode `\$12\catcode
  `\&12\catcode `\#12\catcode `\^12\catcode `\_12\catcode `\%12\relax}%
\providecommand \@@startlink[1]{}%
\providecommand \@@endlink[0]{}%
\providecommand \url  [0]{\begingroup\@sanitize@url \@url }%
\providecommand \@url [1]{\endgroup\@href {#1}{\urlprefix }}%
\providecommand \urlprefix  [0]{URL }%
\providecommand \Eprint [0]{\href }%
\providecommand \doibase [0]{http://dx.doi.org/}%
\providecommand \selectlanguage [0]{\@gobble}%
\providecommand \bibinfo  [0]{\@secondoftwo}%
\providecommand \bibfield  [0]{\@secondoftwo}%
\providecommand \translation [1]{[#1]}%
\providecommand \BibitemOpen [0]{}%
\providecommand \bibitemStop [0]{}%
\providecommand \bibitemNoStop [0]{.\EOS\space}%
\providecommand \EOS [0]{\spacefactor3000\relax}%
\providecommand \BibitemShut  [1]{\csname bibitem#1\endcsname}%
\let\auto@bib@innerbib\@empty
\bibitem [{\citenamefont
  {Glauber}(1963{\natexlab{a}})}]{Glauber1963_CoherentStates}%
  \BibitemOpen
  \bibfield  {author} {\bibinfo {author} {\bibfnamefont {R.~J.}\ \bibnamefont
  {Glauber}},\ }\href@noop {} {\bibfield  {journal} {\bibinfo  {journal} {Phys.
  Rev.}\ }\textbf {\bibinfo {volume} {130}},\ \bibinfo {pages} {2529} (\bibinfo
  {year} {1963}{\natexlab{a}})}\BibitemShut {NoStop}%
\bibitem [{\citenamefont {Glauber}(1963{\natexlab{b}})}]{Glauber_1963_P-Rep}%
  \BibitemOpen
  \bibfield  {author} {\bibinfo {author} {\bibfnamefont {R.~J.}\ \bibnamefont
  {Glauber}},\ }\href@noop {} {\bibfield  {journal} {\bibinfo  {journal} {Phys.
  Rev.}\ }\textbf {\bibinfo {volume} {131}},\ \bibinfo {pages} {2766} (\bibinfo
  {year} {1963}{\natexlab{b}})}\BibitemShut {NoStop}%
\bibitem [{\citenamefont {Aaronson}\ and\ \citenamefont
  {Arkhipov}(2011)}]{AaronsonArkhipov:2011}%
  \BibitemOpen
  \bibfield  {author} {\bibinfo {author} {\bibfnamefont {S.}~\bibnamefont
  {Aaronson}}\ and\ \bibinfo {author} {\bibfnamefont {A.}~\bibnamefont
  {Arkhipov}},\ }in\ \href@noop {} {\emph {\bibinfo {booktitle} {Proceedings of
  the 43rd Annual ACM Symposium on Theory of Computing}}}\ (\bibinfo {address}
  {ACM Press},\ \bibinfo {year} {2011})\ pp.\ \bibinfo {pages}
  {333--342}\BibitemShut {NoStop}%
\bibitem [{\citenamefont {Aaronson}\ and\ \citenamefont
  {Arkhipov}(2013)}]{AaronsonArkhipov2013LV}%
  \BibitemOpen
  \bibfield  {author} {\bibinfo {author} {\bibfnamefont {S.}~\bibnamefont
  {Aaronson}}\ and\ \bibinfo {author} {\bibfnamefont {A.}~\bibnamefont
  {Arkhipov}},\ }\href@noop {} {\bibfield  {journal} {\bibinfo  {journal}
  {Theory of Computing}\ }\textbf {\bibinfo {volume} {9}},\ \bibinfo {pages}
  {143} (\bibinfo {year} {2013})}\BibitemShut {NoStop}%
\bibitem [{\citenamefont {Motes}\ \emph {et~al.}(2015)\citenamefont {Motes}
  \emph {et~al.}}]{Motes2015_PRL114}%
  \BibitemOpen
  \bibfield  {author} {\bibinfo {author} {\bibfnamefont {K.~R.}\ \bibnamefont
  {Motes}} \emph {et~al.},\ }\href@noop {} {\bibfield  {journal} {\bibinfo
  {journal} {Phys. Rev. Lett.}\ }\textbf {\bibinfo {volume} {114}},\ \bibinfo
  {pages} {170802} (\bibinfo {year} {2015})}\BibitemShut {NoStop}%
\bibitem [{\citenamefont {Drummond}\ and\ \citenamefont
  {Gardiner}(1980)}]{Drummond_Gardiner_PositivePRep}%
  \BibitemOpen
  \bibfield  {author} {\bibinfo {author} {\bibfnamefont {P.~D.}\ \bibnamefont
  {Drummond}}\ and\ \bibinfo {author} {\bibfnamefont {C.~W.}\ \bibnamefont
  {Gardiner}},\ }\href@noop {} {\bibfield  {journal} {\bibinfo  {journal} {J.
  Phys. A}\ }\textbf {\bibinfo {volume} {13}},\ \bibinfo {pages} {2353}
  (\bibinfo {year} {1980})}\BibitemShut {NoStop}%
\bibitem [{\citenamefont {Carolan}\ \emph {et~al.}(2015)\citenamefont {Carolan}
  \emph {et~al.}}]{Carolan2015}%
  \BibitemOpen
  \bibfield  {author} {\bibinfo {author} {\bibfnamefont {J.}~\bibnamefont
  {Carolan}} \emph {et~al.},\ }\href@noop {} {\bibfield  {journal} {\bibinfo
  {journal} {Science}\ }\textbf {\bibinfo {volume} {349}},\ \bibinfo {pages}
  {711} (\bibinfo {year} {2015})}\BibitemShut {NoStop}%
\bibitem [{\citenamefont {Aaronson}(2011)}]{Aaronson2011}%
  \BibitemOpen
  \bibfield  {author} {\bibinfo {author} {\bibfnamefont {S.}~\bibnamefont
  {Aaronson}},\ }\href@noop {} {\bibfield  {journal} {\bibinfo  {journal}
  {Proceedings of the Royal Society of London A: Mathematical, Physical and
  Engineering Sciences}\ }\textbf {\bibinfo {volume} {467}},\ \bibinfo {pages}
  {3393} (\bibinfo {year} {2011})}\BibitemShut {NoStop}%
\bibitem [{\citenamefont {Clifford}\ and\ \citenamefont
  {Clifford}(2017)}]{Clifford2017-classical}%
  \BibitemOpen
  \bibfield  {author} {\bibinfo {author} {\bibfnamefont {P.}~\bibnamefont
  {Clifford}}\ and\ \bibinfo {author} {\bibfnamefont {R.}~\bibnamefont
  {Clifford}},\ }\href@noop {} {\bibfield  {journal} {\bibinfo  {journal}
  {arXiv:1706.01260}\ } (\bibinfo {year} {2017})}\BibitemShut {NoStop}%
\bibitem [{\citenamefont {Su}\ \emph {et~al.}(2017)\citenamefont {Su},
  \citenamefont {Li}, \citenamefont {Rohde}, \citenamefont {Huang},
  \citenamefont {Wang}, \citenamefont {Li}, \citenamefont {Liu}, \citenamefont
  {Dowling}, \citenamefont {Lu},\ and\ \citenamefont {Pan}}]{Su:2017}%
  \BibitemOpen
  \bibfield  {author} {\bibinfo {author} {\bibfnamefont {Z.-E.}\ \bibnamefont
  {Su}}, \bibinfo {author} {\bibfnamefont {Y.}~\bibnamefont {Li}}, \bibinfo
  {author} {\bibfnamefont {P.~P.}\ \bibnamefont {Rohde}}, \bibinfo {author}
  {\bibfnamefont {H.-L.}\ \bibnamefont {Huang}}, \bibinfo {author}
  {\bibfnamefont {X.-L.}\ \bibnamefont {Wang}}, \bibinfo {author}
  {\bibfnamefont {L.}~\bibnamefont {Li}}, \bibinfo {author} {\bibfnamefont
  {N.-L.}\ \bibnamefont {Liu}}, \bibinfo {author} {\bibfnamefont {J.~P.}\
  \bibnamefont {Dowling}}, \bibinfo {author} {\bibfnamefont {C.-Y.}\
  \bibnamefont {Lu}}, \ and\ \bibinfo {author} {\bibfnamefont {J.-W.}\
  \bibnamefont {Pan}},\ }\href@noop {} {\bibfield  {journal} {\bibinfo
  {journal} {Phys. Rev. Lett.}\ }\textbf {\bibinfo {volume} {119}},\ \bibinfo
  {pages} {080502} (\bibinfo {year} {2017})}\BibitemShut {NoStop}%
\bibitem [{\citenamefont {Broome}\ \emph {et~al.}(2013)\citenamefont {Broome}
  \emph {et~al.}}]{Broome2013}%
  \BibitemOpen
  \bibfield  {author} {\bibinfo {author} {\bibfnamefont {M.~A.}\ \bibnamefont
  {Broome}} \emph {et~al.},\ }\href@noop {} {\bibfield  {journal} {\bibinfo
  {journal} {Science}\ }\textbf {\bibinfo {volume} {339}},\ \bibinfo {pages}
  {794} (\bibinfo {year} {2013})}\BibitemShut {NoStop}%
\bibitem [{\citenamefont {Crespi}\ \emph {et~al.}(2013)\citenamefont {Crespi}
  \emph {et~al.}}]{Crespi2013}%
  \BibitemOpen
  \bibfield  {author} {\bibinfo {author} {\bibfnamefont {A.}~\bibnamefont
  {Crespi}} \emph {et~al.},\ }\href@noop {} {\bibfield  {journal} {\bibinfo
  {journal} {Nat. Photon.}\ }\textbf {\bibinfo {volume} {7}},\ \bibinfo {pages}
  {545} (\bibinfo {year} {2013})}\BibitemShut {NoStop}%
\bibitem [{\citenamefont {Tillmann}\ \emph {et~al.}(2013)\citenamefont
  {Tillmann} \emph {et~al.}}]{Tillmann2013}%
  \BibitemOpen
  \bibfield  {author} {\bibinfo {author} {\bibfnamefont {M.}~\bibnamefont
  {Tillmann}} \emph {et~al.},\ }\href@noop {} {\bibfield  {journal} {\bibinfo
  {journal} {Nat. Photon.}\ }\textbf {\bibinfo {volume} {7}},\ \bibinfo {pages}
  {540} (\bibinfo {year} {2013})}\BibitemShut {NoStop}%
\bibitem [{\citenamefont {Spring}\ \emph {et~al.}(2013)\citenamefont {Spring}
  \emph {et~al.}}]{Spring72013}%
  \BibitemOpen
  \bibfield  {author} {\bibinfo {author} {\bibfnamefont {J.~B.}\ \bibnamefont
  {Spring}} \emph {et~al.},\ }\href@noop {} {\bibfield  {journal} {\bibinfo
  {journal} {Science}\ }\textbf {\bibinfo {volume} {339}},\ \bibinfo {pages}
  {798} (\bibinfo {year} {2013})}\BibitemShut {NoStop}%
\bibitem [{\citenamefont {Crespi}\ \emph {et~al.}(2016)\citenamefont {Crespi}
  \emph {et~al.}}]{Crespi:2014}%
  \BibitemOpen
  \bibfield  {author} {\bibinfo {author} {\bibfnamefont {A.}~\bibnamefont
  {Crespi}} \emph {et~al.},\ }\href@noop {} {\bibfield  {journal} {\bibinfo
  {journal} {Nat. Commun.}\ }\textbf {\bibinfo {volume} {7}},\ \bibinfo {pages}
  {10469} (\bibinfo {year} {2016})}\BibitemShut {NoStop}%
\bibitem [{\citenamefont {Spagnolo}\ \emph {et~al.}(2014)\citenamefont
  {Spagnolo} \emph {et~al.}}]{Spagnolo2014}%
  \BibitemOpen
  \bibfield  {author} {\bibinfo {author} {\bibfnamefont {N.}~\bibnamefont
  {Spagnolo}} \emph {et~al.},\ }\href@noop {} {\bibfield  {journal} {\bibinfo
  {journal} {Nat. Photon.}\ }\textbf {\bibinfo {volume} {8}},\ \bibinfo {pages}
  {615} (\bibinfo {year} {2014})}\BibitemShut {NoStop}%
\bibitem [{\citenamefont {Wang}\ \emph {et~al.}(2017)\citenamefont {Wang} \emph
  {et~al.}}]{Wang2017}%
  \BibitemOpen
  \bibfield  {author} {\bibinfo {author} {\bibfnamefont {H.}~\bibnamefont
  {Wang}} \emph {et~al.},\ }\href@noop {} {\bibfield  {journal} {\bibinfo
  {journal} {Nat. Photon.}\ }\textbf {\bibinfo {volume} {11}},\ \bibinfo
  {pages} {361} (\bibinfo {year} {2017})}\BibitemShut {NoStop}%
\bibitem [{\citenamefont {Loredo}\ \emph {et~al.}(2017)\citenamefont {Loredo},
  \citenamefont {Broome}, \citenamefont {Hilaire}, \citenamefont {Gazzano},
  \citenamefont {Sagnes}, \citenamefont {Lemaitre}, \citenamefont {Almeida},
  \citenamefont {Senellart},\ and\ \citenamefont {White}}]{Loredo2016}%
  \BibitemOpen
  \bibfield  {author} {\bibinfo {author} {\bibfnamefont {J.~C.}\ \bibnamefont
  {Loredo}}, \bibinfo {author} {\bibfnamefont {M.~A.}\ \bibnamefont {Broome}},
  \bibinfo {author} {\bibfnamefont {P.}~\bibnamefont {Hilaire}}, \bibinfo
  {author} {\bibfnamefont {O.}~\bibnamefont {Gazzano}}, \bibinfo {author}
  {\bibfnamefont {I.}~\bibnamefont {Sagnes}}, \bibinfo {author} {\bibfnamefont
  {A.}~\bibnamefont {Lemaitre}}, \bibinfo {author} {\bibfnamefont {M.~P.}\
  \bibnamefont {Almeida}}, \bibinfo {author} {\bibfnamefont {P.}~\bibnamefont
  {Senellart}}, \ and\ \bibinfo {author} {\bibfnamefont {A.~G.}\ \bibnamefont
  {White}},\ }\href@noop {} {\bibfield  {journal} {\bibinfo  {journal} {Phys.
  Rev. Lett.}\ }\textbf {\bibinfo {volume} {118}},\ \bibinfo {pages} {130503}
  (\bibinfo {year} {2017})}\BibitemShut {NoStop}%
\bibitem [{\citenamefont {Scheel}(2004)}]{Scheel2004Permanents}%
  \BibitemOpen
  \bibfield  {author} {\bibinfo {author} {\bibfnamefont {S.}~\bibnamefont
  {Scheel}},\ }\href@noop {} {\bibfield  {journal} {\bibinfo  {journal}
  {arXiv:quant-ph/0406127}\ } (\bibinfo {year} {2004})}\BibitemShut {NoStop}%
\bibitem [{\citenamefont {Scheel}(2005)}]{Scheel2005}%
  \BibitemOpen
  \bibfield  {author} {\bibinfo {author} {\bibfnamefont {S.}~\bibnamefont
  {Scheel}},\ }in\ \href@noop {} {\emph {\bibinfo {booktitle} {Quantum
  Information Processing}}},\ \bibinfo {editor} {edited by\ \bibinfo {editor}
  {\bibfnamefont {T.}~\bibnamefont {Beth}}\ and\ \bibinfo {editor}
  {\bibfnamefont {G.}~\bibnamefont {Leuchs}}}\ (\bibinfo  {publisher}
  {Wiley-VCH, Weinheim},\ \bibinfo {year} {2005})\ Chap.~\bibinfo {chapter}
  {28}, pp.\ \bibinfo {pages} {382--392}\BibitemShut {NoStop}%
\bibitem [{\citenamefont {Valiant}(1979)}]{Valiant1979}%
  \BibitemOpen
  \bibfield  {author} {\bibinfo {author} {\bibfnamefont {L.}~\bibnamefont
  {Valiant}},\ }\href@noop {} {\bibfield  {journal} {\bibinfo  {journal}
  {Theoretical Computer Science}\ }\textbf {\bibinfo {volume} {8}},\ \bibinfo
  {pages} {189} (\bibinfo {year} {1979})}\BibitemShut {NoStop}%
\bibitem [{\citenamefont {Glynn}(2010)}]{Glynn2010Perm}%
  \BibitemOpen
  \bibfield  {author} {\bibinfo {author} {\bibfnamefont {D.~G.}\ \bibnamefont
  {Glynn}},\ }\href@noop {} {\bibfield  {journal} {\bibinfo  {journal}
  {European Journal of Combinatorics}\ }\textbf {\bibinfo {volume} {31}},\
  \bibinfo {pages} {1887 } (\bibinfo {year} {2010})}\BibitemShut {NoStop}%
\bibitem [{\citenamefont {{Wu}}\ \emph {et~al.}(2016)\citenamefont {{Wu}},
  \citenamefont {{Liu}}, \citenamefont {{Zhang}}, \citenamefont {{Jin}},
  \citenamefont {{Wang}}, \citenamefont {{Wang}},\ and\ \citenamefont
  {{Yang}}}]{Wu2016arXiv160605836}%
  \BibitemOpen
  \bibfield  {author} {\bibinfo {author} {\bibfnamefont {J.}~\bibnamefont
  {{Wu}}}, \bibinfo {author} {\bibfnamefont {Y.}~\bibnamefont {{Liu}}},
  \bibinfo {author} {\bibfnamefont {B.}~\bibnamefont {{Zhang}}}, \bibinfo
  {author} {\bibfnamefont {X.}~\bibnamefont {{Jin}}}, \bibinfo {author}
  {\bibfnamefont {Y.}~\bibnamefont {{Wang}}}, \bibinfo {author} {\bibfnamefont
  {H.}~\bibnamefont {{Wang}}}, \ and\ \bibinfo {author} {\bibfnamefont
  {X.}~\bibnamefont {{Yang}}},\ }\href@noop {} {\bibfield  {journal} {\bibinfo
  {journal} {ArXiv e-prints}\ } (\bibinfo {year} {2016})},\ \Eprint
  {http://arxiv.org/abs/1606.05836} {arXiv:1606.05836} \BibitemShut {NoStop}%
\bibitem [{\citenamefont {Opanchuk}\ \emph {et~al.}(2017)\citenamefont
  {Opanchuk}, \citenamefont {Rosales-Z\'arate}, \citenamefont {Reid},\ and\
  \citenamefont {Drummond}}]{Opanchuk2017-robustness}%
  \BibitemOpen
  \bibfield  {author} {\bibinfo {author} {\bibfnamefont {B.}~\bibnamefont
  {Opanchuk}}, \bibinfo {author} {\bibfnamefont {L.}~\bibnamefont
  {Rosales-Z\'arate}}, \bibinfo {author} {\bibfnamefont {M.~D.}\ \bibnamefont
  {Reid}}, \ and\ \bibinfo {author} {\bibfnamefont {P.~D.}\ \bibnamefont
  {Drummond}},\ }\href@noop {} {\bibfield  {journal} {\bibinfo  {journal}
  {arXiv:1711.07153}\ } (\bibinfo {year} {2017})}\BibitemShut {NoStop}%
\bibitem [{\citenamefont {He}\ \emph {et~al.}(2017)\citenamefont {He} \emph
  {et~al.}}]{He2016ScalableBS}%
  \BibitemOpen
  \bibfield  {author} {\bibinfo {author} {\bibfnamefont {Y.}~\bibnamefont {He}}
  \emph {et~al.},\ }\href@noop {} {\bibfield  {journal} {\bibinfo  {journal}
  {Phys. Rev. Lett.}\ }\textbf {\bibinfo {volume} {118}},\ \bibinfo {pages}
  {190501} (\bibinfo {year} {2017})}\BibitemShut {NoStop}%
\bibitem [{\citenamefont {Olson}\ \emph {et~al.}(2017)\citenamefont {Olson},
  \citenamefont {Motes}, \citenamefont {Birchall}, \citenamefont {Studer},
  \citenamefont {LaBorde}, \citenamefont {Moulder}, \citenamefont {Rohde},\
  and\ \citenamefont {Dowling}}]{OlsonLinearOptQM}%
  \BibitemOpen
  \bibfield  {author} {\bibinfo {author} {\bibfnamefont {J.~P.}\ \bibnamefont
  {Olson}}, \bibinfo {author} {\bibfnamefont {K.~R.}\ \bibnamefont {Motes}},
  \bibinfo {author} {\bibfnamefont {P.~M.}\ \bibnamefont {Birchall}}, \bibinfo
  {author} {\bibfnamefont {N.~M.}\ \bibnamefont {Studer}}, \bibinfo {author}
  {\bibfnamefont {M.}~\bibnamefont {LaBorde}}, \bibinfo {author} {\bibfnamefont
  {T.}~\bibnamefont {Moulder}}, \bibinfo {author} {\bibfnamefont {P.~P.}\
  \bibnamefont {Rohde}}, \ and\ \bibinfo {author} {\bibfnamefont {J.~P.}\
  \bibnamefont {Dowling}},\ }\href@noop {} {\bibfield  {journal} {\bibinfo
  {journal} {Phys. Rev. A}\ }\textbf {\bibinfo {volume} {96}},\ \bibinfo
  {pages} {013810} (\bibinfo {year} {2017})}\BibitemShut {NoStop}%
\bibitem [{\citenamefont {Dirac}(1945)}]{Dirac_RevModPhys_1945}%
  \BibitemOpen
  \bibfield  {author} {\bibinfo {author} {\bibfnamefont {P.~A.~M.}\
  \bibnamefont {Dirac}},\ }\href {\doibase 10.1103/RevModPhys.17.195}
  {\bibfield  {journal} {\bibinfo  {journal} {Rev. Mod. Phys.}\ }\textbf
  {\bibinfo {volume} {17}},\ \bibinfo {pages} {195} (\bibinfo {year}
  {1945})}\BibitemShut {NoStop}%
\bibitem [{\citenamefont {Drummond}\ and\ \citenamefont
  {Hillery}(2014)}]{drummond2014quantum}%
  \BibitemOpen
  \bibfield  {author} {\bibinfo {author} {\bibfnamefont {P.~D.}\ \bibnamefont
  {Drummond}}\ and\ \bibinfo {author} {\bibfnamefont {M.}~\bibnamefont
  {Hillery}},\ }\href@noop {} {\emph {\bibinfo {title} {The Quantum Theory of
  Nonlinear Optics}}}\ (\bibinfo  {publisher} {Cambridge University Press},\
  \bibinfo {year} {2014})\BibitemShut {NoStop}%
\bibitem [{\citenamefont {Tichy}\ \emph {et~al.}(2014)\citenamefont {Tichy},
  \citenamefont {Mayer}, \citenamefont {Buchleitner},\ and\ \citenamefont
  {M\o{}lmer}}]{Tichy2014}%
  \BibitemOpen
  \bibfield  {author} {\bibinfo {author} {\bibfnamefont {M.~C.}\ \bibnamefont
  {Tichy}}, \bibinfo {author} {\bibfnamefont {K.}~\bibnamefont {Mayer}},
  \bibinfo {author} {\bibfnamefont {A.}~\bibnamefont {Buchleitner}}, \ and\
  \bibinfo {author} {\bibfnamefont {K.}~\bibnamefont {M\o{}lmer}},\ }\href@noop
  {} {\bibfield  {journal} {\bibinfo  {journal} {Phys. Rev. Lett.}\ }\textbf
  {\bibinfo {volume} {113}},\ \bibinfo {pages} {020502} (\bibinfo {year}
  {2014})}\BibitemShut {NoStop}%
\bibitem [{\citenamefont {Mayer}\ \emph {et~al.}(2011)\citenamefont {Mayer},
  \citenamefont {Tichy}, \citenamefont {Mintert}, \citenamefont {Konrad},\ and\
  \citenamefont {Buchleitner}}]{Mayer2011}%
  \BibitemOpen
  \bibfield  {author} {\bibinfo {author} {\bibfnamefont {K.}~\bibnamefont
  {Mayer}}, \bibinfo {author} {\bibfnamefont {M.~C.}\ \bibnamefont {Tichy}},
  \bibinfo {author} {\bibfnamefont {F.}~\bibnamefont {Mintert}}, \bibinfo
  {author} {\bibfnamefont {T.}~\bibnamefont {Konrad}}, \ and\ \bibinfo {author}
  {\bibfnamefont {A.}~\bibnamefont {Buchleitner}},\ }\href@noop {} {\bibfield
  {journal} {\bibinfo  {journal} {Phys. Rev. A}\ }\textbf {\bibinfo {volume}
  {83}},\ \bibinfo {pages} {062307} (\bibinfo {year} {2011})}\BibitemShut
  {NoStop}%
\bibitem [{\citenamefont {Walschaers}\ \emph {et~al.}(2016)\citenamefont
  {Walschaers} \emph {et~al.}}]{Walschaers2016_NJP18}%
  \BibitemOpen
  \bibfield  {author} {\bibinfo {author} {\bibfnamefont {M.}~\bibnamefont
  {Walschaers}} \emph {et~al.},\ }\href@noop {} {\bibfield  {journal} {\bibinfo
   {journal} {New Journal of Physics}\ }\textbf {\bibinfo {volume} {18}},\
  \bibinfo {pages} {032001} (\bibinfo {year} {2016})}\BibitemShut {NoStop}%
\bibitem [{\citenamefont {Drummond}\ and\ \citenamefont
  {Walls}(1980)}]{drummond1980quantum}%
  \BibitemOpen
  \bibfield  {author} {\bibinfo {author} {\bibfnamefont {P.}~\bibnamefont
  {Drummond}}\ and\ \bibinfo {author} {\bibfnamefont {D.}~\bibnamefont
  {Walls}},\ }\href@noop {} {\bibfield  {journal} {\bibinfo  {journal} {Journal
  of Physics A: Mathematical and General}\ }\textbf {\bibinfo {volume} {13}},\
  \bibinfo {pages} {725} (\bibinfo {year} {1980})}\BibitemShut {NoStop}%
\bibitem [{\citenamefont {Bartolo}\ \emph {et~al.}(2016)\citenamefont
  {Bartolo}, \citenamefont {Minganti}, \citenamefont {Casteels},\ and\
  \citenamefont {Ciuti}}]{Bartolo:2016}%
  \BibitemOpen
  \bibfield  {author} {\bibinfo {author} {\bibfnamefont {N.}~\bibnamefont
  {Bartolo}}, \bibinfo {author} {\bibfnamefont {F.}~\bibnamefont {Minganti}},
  \bibinfo {author} {\bibfnamefont {W.}~\bibnamefont {Casteels}}, \ and\
  \bibinfo {author} {\bibfnamefont {C.}~\bibnamefont {Ciuti}},\ }\href@noop {}
  {\bibfield  {journal} {\bibinfo  {journal} {Phys. Rev. A}\ }\textbf {\bibinfo
  {volume} {94}},\ \bibinfo {pages} {033841} (\bibinfo {year}
  {2016})}\BibitemShut {NoStop}%
\bibitem [{\citenamefont {Minganti}\ \emph {et~al.}(2016)\citenamefont
  {Minganti}, \citenamefont {Bartolo}, \citenamefont {Lolli}, \citenamefont
  {Casteels},\ and\ \citenamefont {Ciuti}}]{Minganti:2016}%
  \BibitemOpen
  \bibfield  {author} {\bibinfo {author} {\bibfnamefont {F.}~\bibnamefont
  {Minganti}}, \bibinfo {author} {\bibfnamefont {N.}~\bibnamefont {Bartolo}},
  \bibinfo {author} {\bibfnamefont {J.}~\bibnamefont {Lolli}}, \bibinfo
  {author} {\bibfnamefont {W.}~\bibnamefont {Casteels}}, \ and\ \bibinfo
  {author} {\bibfnamefont {C.}~\bibnamefont {Ciuti}},\ }\href@noop {}
  {\bibfield  {journal} {\bibinfo  {journal} {Scientific Reports}\ }\textbf
  {\bibinfo {volume} {6}},\ \bibinfo {pages} {26987} (\bibinfo {year}
  {2016})}\BibitemShut {NoStop}%
\bibitem [{\citenamefont {Drummond}\ \emph
  {et~al.}(1981{\natexlab{a}})\citenamefont {Drummond}, \citenamefont
  {Gardiner},\ and\ \citenamefont {Walls}}]{DrummondGardinerWalls1981}%
  \BibitemOpen
  \bibfield  {author} {\bibinfo {author} {\bibfnamefont {P.~D.}\ \bibnamefont
  {Drummond}}, \bibinfo {author} {\bibfnamefont {C.~W.}\ \bibnamefont
  {Gardiner}}, \ and\ \bibinfo {author} {\bibfnamefont {D.~F.}\ \bibnamefont
  {Walls}},\ }\href@noop {} {\bibfield  {journal} {\bibinfo  {journal} {Phys.
  Rev. A}\ }\textbf {\bibinfo {volume} {24}},\ \bibinfo {pages} {914} (\bibinfo
  {year} {1981}{\natexlab{a}})}\BibitemShut {NoStop}%
\bibitem [{\citenamefont {Drummond}\ \emph
  {et~al.}(1981{\natexlab{b}})\citenamefont {Drummond}, \citenamefont
  {McNeil},\ and\ \citenamefont
  {Walls}}]{drummond1981_II_nonequilibriumparamp}%
  \BibitemOpen
  \bibfield  {author} {\bibinfo {author} {\bibfnamefont {P.}~\bibnamefont
  {Drummond}}, \bibinfo {author} {\bibfnamefont {K.}~\bibnamefont {McNeil}}, \
  and\ \bibinfo {author} {\bibfnamefont {D.}~\bibnamefont {Walls}},\
  }\href@noop {} {\bibfield  {journal} {\bibinfo  {journal} {J. Mod. Opt.}\
  }\textbf {\bibinfo {volume} {28}},\ \bibinfo {pages} {211} (\bibinfo {year}
  {1981}{\natexlab{b}})}\BibitemShut {NoStop}%
\bibitem [{\citenamefont {McNeil}\ and\ \citenamefont
  {Gardiner}(1983)}]{mcneil1983quantum}%
  \BibitemOpen
  \bibfield  {author} {\bibinfo {author} {\bibfnamefont {K.}~\bibnamefont
  {McNeil}}\ and\ \bibinfo {author} {\bibfnamefont {C.}~\bibnamefont
  {Gardiner}},\ }\href@noop {} {\bibfield  {journal} {\bibinfo  {journal}
  {Physical Review A}\ }\textbf {\bibinfo {volume} {28}},\ \bibinfo {pages}
  {1560} (\bibinfo {year} {1983})}\BibitemShut {NoStop}%
\bibitem [{\citenamefont {Cao}\ \emph {et~al.}(2016)\citenamefont {Cao},
  \citenamefont {Mahmud},\ and\ \citenamefont {Hafezi}}]{Cao:2016}%
  \BibitemOpen
  \bibfield  {author} {\bibinfo {author} {\bibfnamefont {B.}~\bibnamefont
  {Cao}}, \bibinfo {author} {\bibfnamefont {K.~W.}\ \bibnamefont {Mahmud}}, \
  and\ \bibinfo {author} {\bibfnamefont {M.}~\bibnamefont {Hafezi}},\
  }\href@noop {} {\bibfield  {journal} {\bibinfo  {journal} {Phys. Rev. A}\
  }\textbf {\bibinfo {volume} {94}},\ \bibinfo {pages} {063805} (\bibinfo
  {year} {2016})}\BibitemShut {NoStop}%
\bibitem [{\citenamefont {Drummond}(2017)}]{drummond2016coherent}%
  \BibitemOpen
  \bibfield  {author} {\bibinfo {author} {\bibfnamefont {P.}~\bibnamefont
  {Drummond}},\ }\href@noop {} {\bibfield  {journal} {\bibinfo  {journal} {J.
  Phys. A}\ }\textbf {\bibinfo {volume} {50}},\ \bibinfo {pages} {45LT01}
  (\bibinfo {year} {2017})}\BibitemShut {NoStop}%
\bibitem [{\citenamefont {Drummond}\ and\ \citenamefont
  {Reid}(2016)}]{Drummond2016-coherent-states}%
  \BibitemOpen
  \bibfield  {author} {\bibinfo {author} {\bibfnamefont {P.~D.}\ \bibnamefont
  {Drummond}}\ and\ \bibinfo {author} {\bibfnamefont {M.~D.}\ \bibnamefont
  {Reid}},\ }\href@noop {} {\bibfield  {journal} {\bibinfo  {journal} {Phys.
  Rev. A}\ }\textbf {\bibinfo {volume} {94}},\ \bibinfo {pages} {063851}
  (\bibinfo {year} {2016})}\BibitemShut {NoStop}%
\bibitem [{\citenamefont {Gurvits}(2005)}]{Gurvits2005}%
  \BibitemOpen
  \bibfield  {author} {\bibinfo {author} {\bibfnamefont {L.}~\bibnamefont
  {Gurvits}},\ }in\ \href@noop {} {\emph {\bibinfo {booktitle} {Mathematical
  Foundations of Computer Science 2005}}},\ \bibinfo {editor} {edited by\
  \bibinfo {editor} {\bibfnamefont {J.}~\bibnamefont {J\k{e}drzejowicz}}\ and\
  \bibinfo {editor} {\bibfnamefont {A.}~\bibnamefont {Szepietowski}}}\
  (\bibinfo  {publisher} {Springer},\ \bibinfo {address} {Berlin, Heidelberg},\
  \bibinfo {year} {2005})\ pp.\ \bibinfo {pages} {447--458}\BibitemShut
  {NoStop}%
\bibitem [{\citenamefont {Perlmutter}\ \emph {et~al.}(1988)\citenamefont
  {Perlmutter}, \citenamefont {Levenson}, \citenamefont {Shelby},\ and\
  \citenamefont {Weissman}}]{perlmutter1988inverse}%
  \BibitemOpen
  \bibfield  {author} {\bibinfo {author} {\bibfnamefont {S.}~\bibnamefont
  {Perlmutter}}, \bibinfo {author} {\bibfnamefont {M.}~\bibnamefont
  {Levenson}}, \bibinfo {author} {\bibfnamefont {R.}~\bibnamefont {Shelby}}, \
  and\ \bibinfo {author} {\bibfnamefont {M.}~\bibnamefont {Weissman}},\
  }\href@noop {} {\bibfield  {journal} {\bibinfo  {journal} {Physical review
  letters}\ }\textbf {\bibinfo {volume} {61}},\ \bibinfo {pages} {1388}
  (\bibinfo {year} {1988})}\BibitemShut {NoStop}%
\bibitem [{\citenamefont {Shelby}\ \emph {et~al.}(1990)\citenamefont {Shelby},
  \citenamefont {Drummond},\ and\ \citenamefont {Carter}}]{shelby1990phase}%
  \BibitemOpen
  \bibfield  {author} {\bibinfo {author} {\bibfnamefont {R.}~\bibnamefont
  {Shelby}}, \bibinfo {author} {\bibfnamefont {P.}~\bibnamefont {Drummond}}, \
  and\ \bibinfo {author} {\bibfnamefont {S.}~\bibnamefont {Carter}},\
  }\href@noop {} {\bibfield  {journal} {\bibinfo  {journal} {Physical Review
  A}\ }\textbf {\bibinfo {volume} {42}},\ \bibinfo {pages} {2966} (\bibinfo
  {year} {1990})}\BibitemShut {NoStop}%
\bibitem [{\citenamefont {Carter}\ and\ \citenamefont
  {Drummond}(1991)}]{Carter:1991}%
  \BibitemOpen
  \bibfield  {author} {\bibinfo {author} {\bibfnamefont {S.~J.}\ \bibnamefont
  {Carter}}\ and\ \bibinfo {author} {\bibfnamefont {P.~D.}\ \bibnamefont
  {Drummond}},\ }\href@noop {} {\bibfield  {journal} {\bibinfo  {journal}
  {Phys. Rev. Lett.}\ }\textbf {\bibinfo {volume} {67}},\ \bibinfo {pages}
  {3757} (\bibinfo {year} {1991})}\BibitemShut {NoStop}%
\bibitem [{\citenamefont {Drummond}\ and\ \citenamefont
  {Corney}(2001)}]{Drummond:2001a}%
  \BibitemOpen
  \bibfield  {author} {\bibinfo {author} {\bibfnamefont {P.~D.}\ \bibnamefont
  {Drummond}}\ and\ \bibinfo {author} {\bibfnamefont {J.~F.}\ \bibnamefont
  {Corney}},\ }\href@noop {} {\bibfield  {journal} {\bibinfo  {journal} {J.
  Opt. Soc. Am. B}\ }\textbf {\bibinfo {volume} {18}},\ \bibinfo {pages} {139}
  (\bibinfo {year} {2001})}\BibitemShut {NoStop}%
\bibitem [{\citenamefont {Arkhipov}\ and\ \citenamefont
  {Kuperberg}(2012)}]{Arkhipov2012bBsonicBP}%
  \BibitemOpen
  \bibfield  {author} {\bibinfo {author} {\bibfnamefont {A.}~\bibnamefont
  {Arkhipov}}\ and\ \bibinfo {author} {\bibfnamefont {G.}~\bibnamefont
  {Kuperberg}},\ }\href@noop {} {\bibfield  {journal} {\bibinfo  {journal}
  {Geometry \& Topology Monographs}\ }\textbf {\bibinfo {volume} {18}},\
  \bibinfo {pages} {1} (\bibinfo {year} {2012})}\BibitemShut {NoStop}%
\bibitem [{\citenamefont {Drummond}\ \emph {et~al.}(2016)\citenamefont
  {Drummond}, \citenamefont {Opanchuk}, \citenamefont {Rosales-Z\'arate},
  \citenamefont {Reid},\ and\ \citenamefont {Forrester}}]{ScalingPerm}%
  \BibitemOpen
  \bibfield  {author} {\bibinfo {author} {\bibfnamefont {P.~D.}\ \bibnamefont
  {Drummond}}, \bibinfo {author} {\bibfnamefont {B.}~\bibnamefont {Opanchuk}},
  \bibinfo {author} {\bibfnamefont {L.}~\bibnamefont {Rosales-Z\'arate}},
  \bibinfo {author} {\bibfnamefont {M.~D.}\ \bibnamefont {Reid}}, \ and\
  \bibinfo {author} {\bibfnamefont {P.~J.}\ \bibnamefont {Forrester}},\
  }\href@noop {} {\bibfield  {journal} {\bibinfo  {journal} {Phys. Rev. A}\
  }\textbf {\bibinfo {volume} {94}},\ \bibinfo {pages} {042339} (\bibinfo
  {year} {2016})}\BibitemShut {NoStop}%
\bibitem [{\citenamefont {Gurvits}\ and\ \citenamefont
  {Samorodnitsky}(2002)}]{gurvits2002deterministic}%
  \BibitemOpen
  \bibfield  {author} {\bibinfo {author} {\bibfnamefont {L.}~\bibnamefont
  {Gurvits}}\ and\ \bibinfo {author} {\bibfnamefont {A.}~\bibnamefont
  {Samorodnitsky}},\ }\href@noop {} {\bibfield  {journal} {\bibinfo  {journal}
  {Discrete \& Computational Geometry}\ }\textbf {\bibinfo {volume} {27}},\
  \bibinfo {pages} {531} (\bibinfo {year} {2002})}\BibitemShut {NoStop}%
\bibitem [{\citenamefont {Carolan}\ \emph {et~al.}(2014)\citenamefont {Carolan}
  \emph {et~al.}}]{Carolan2014}%
  \BibitemOpen
  \bibfield  {author} {\bibinfo {author} {\bibfnamefont {J.}~\bibnamefont
  {Carolan}} \emph {et~al.},\ }\href@noop {} {\bibfield  {journal} {\bibinfo
  {journal} {Nat. Photon.}\ }\textbf {\bibinfo {volume} {8}},\ \bibinfo {pages}
  {621} (\bibinfo {year} {2014})}\BibitemShut {NoStop}%
\bibitem [{\citenamefont {Aaronson}\ and\ \citenamefont
  {Arkhipov}(2014)}]{Aaronson:2014}%
  \BibitemOpen
  \bibfield  {author} {\bibinfo {author} {\bibfnamefont {S.}~\bibnamefont
  {Aaronson}}\ and\ \bibinfo {author} {\bibfnamefont {A.}~\bibnamefont
  {Arkhipov}},\ }\href@noop {} {\bibfield  {journal} {\bibinfo  {journal}
  {Quantum Info. Comput.}\ }\textbf {\bibinfo {volume} {14}},\ \bibinfo {pages}
  {1383} (\bibinfo {year} {2014})}\BibitemShut {NoStop}%
\bibitem [{\citenamefont {Bentivegna}\ \emph {et~al.}(2015)\citenamefont
  {Bentivegna} \emph {et~al.}}]{Bentivegna2015}%
  \BibitemOpen
  \bibfield  {author} {\bibinfo {author} {\bibfnamefont {M.}~\bibnamefont
  {Bentivegna}} \emph {et~al.},\ }\href@noop {} {\bibfield  {journal} {\bibinfo
   {journal} {Science Advances}\ }\textbf {\bibinfo {volume} {1}},\ \bibinfo
  {pages} {e1400255} (\bibinfo {year} {2015})}\BibitemShut {NoStop}%
\bibitem [{\citenamefont {Aolita}\ \emph {et~al.}(2015)\citenamefont {Aolita},
  \citenamefont {Gogolin}, \citenamefont {Kliesch},\ and\ \citenamefont
  {Eisert}}]{AolitaEisert2015}%
  \BibitemOpen
  \bibfield  {author} {\bibinfo {author} {\bibfnamefont {L.}~\bibnamefont
  {Aolita}}, \bibinfo {author} {\bibfnamefont {C.}~\bibnamefont {Gogolin}},
  \bibinfo {author} {\bibfnamefont {M.}~\bibnamefont {Kliesch}}, \ and\
  \bibinfo {author} {\bibfnamefont {J.}~\bibnamefont {Eisert}},\ }\href@noop {}
  {\bibfield  {journal} {\bibinfo  {journal} {Nat. Commun.}\ }\textbf {\bibinfo
  {volume} {6}},\ \bibinfo {pages} {8498} (\bibinfo {year} {2015})}\BibitemShut
  {NoStop}%
\bibitem [{\citenamefont {Moran}(1968)}]{Moran}%
  \BibitemOpen
  \bibfield  {author} {\bibinfo {author} {\bibfnamefont {P.~A.~P.}\
  \bibnamefont {Moran}},\ }\href@noop {} {\emph {\bibinfo {title} {An
  introduction to Probability Theory}}}\ (\bibinfo  {publisher} {Oxford,
  Clarendon Press},\ \bibinfo {year} {1968})\BibitemShut {NoStop}%
\bibitem [{\citenamefont {Shchesnovich}(2016)}]{Shchesnovich:2016_PRL}%
  \BibitemOpen
  \bibfield  {author} {\bibinfo {author} {\bibfnamefont {V.~S.}\ \bibnamefont
  {Shchesnovich}},\ }\href@noop {} {\bibfield  {journal} {\bibinfo  {journal}
  {Phys. Rev. Lett.}\ }\textbf {\bibinfo {volume} {116}},\ \bibinfo {pages}
  {123601} (\bibinfo {year} {2016})}\BibitemShut {NoStop}%
\bibitem [{\citenamefont {Agresti}\ \emph {et~al.}(2017)\citenamefont
  {Agresti}, \citenamefont {Viggianiello}, \citenamefont {Flamini},
  \citenamefont {Spagnolo}, \citenamefont {Crespi}, \citenamefont {Osellame},
  \citenamefont {Wiebe},\ and\ \citenamefont {Sciarrino}}]{Agresti:2017}%
  \BibitemOpen
  \bibfield  {author} {\bibinfo {author} {\bibfnamefont {I.}~\bibnamefont
  {Agresti}}, \bibinfo {author} {\bibfnamefont {N.}~\bibnamefont
  {Viggianiello}}, \bibinfo {author} {\bibfnamefont {F.}~\bibnamefont
  {Flamini}}, \bibinfo {author} {\bibfnamefont {N.}~\bibnamefont {Spagnolo}},
  \bibinfo {author} {\bibfnamefont {A.}~\bibnamefont {Crespi}}, \bibinfo
  {author} {\bibfnamefont {R.}~\bibnamefont {Osellame}}, \bibinfo {author}
  {\bibfnamefont {N.}~\bibnamefont {Wiebe}}, \ and\ \bibinfo {author}
  {\bibfnamefont {F.}~\bibnamefont {Sciarrino}},\ }\href@noop {} {\bibfield
  {journal} {\bibinfo  {journal} {arXiv:1712.06863}\ } (\bibinfo {year}
  {2017})}\BibitemShut {NoStop}%
\bibitem [{\citenamefont {Viggianiello}\ \emph {et~al.}(2017)\citenamefont
  {Viggianiello}, \citenamefont {Flamini}, \citenamefont {Bentivegna},
  \citenamefont {Spagnolo}, \citenamefont {Crespi}, \citenamefont {Brod},
  \citenamefont {Galv\~ao}, \citenamefont {Osellame},\ and\ \citenamefont
  {Sciarrino}}]{Viggianiello:2017}%
  \BibitemOpen
  \bibfield  {author} {\bibinfo {author} {\bibfnamefont {N.}~\bibnamefont
  {Viggianiello}}, \bibinfo {author} {\bibfnamefont {F.}~\bibnamefont
  {Flamini}}, \bibinfo {author} {\bibfnamefont {M.}~\bibnamefont {Bentivegna}},
  \bibinfo {author} {\bibfnamefont {N.}~\bibnamefont {Spagnolo}}, \bibinfo
  {author} {\bibfnamefont {A.}~\bibnamefont {Crespi}}, \bibinfo {author}
  {\bibfnamefont {D.~J.}\ \bibnamefont {Brod}}, \bibinfo {author}
  {\bibfnamefont {E.~F.}\ \bibnamefont {Galv\~ao}}, \bibinfo {author}
  {\bibfnamefont {R.}~\bibnamefont {Osellame}}, \ and\ \bibinfo {author}
  {\bibfnamefont {F.}~\bibnamefont {Sciarrino}},\ }\href@noop {} {\bibfield
  {journal} {\bibinfo  {journal} {arXiv:1710.03578v2}\ } (\bibinfo {year}
  {2017})}\BibitemShut {NoStop}%
\bibitem [{\citenamefont {Tichy}\ \emph {et~al.}(2010)\citenamefont {Tichy},
  \citenamefont {Tiersch}, \citenamefont {de~Melo}, \citenamefont {Mintert},\
  and\ \citenamefont {Buchleitner}}]{Tichy2010}%
  \BibitemOpen
  \bibfield  {author} {\bibinfo {author} {\bibfnamefont {M.~C.}\ \bibnamefont
  {Tichy}}, \bibinfo {author} {\bibfnamefont {M.}~\bibnamefont {Tiersch}},
  \bibinfo {author} {\bibfnamefont {F.}~\bibnamefont {de~Melo}}, \bibinfo
  {author} {\bibfnamefont {F.}~\bibnamefont {Mintert}}, \ and\ \bibinfo
  {author} {\bibfnamefont {A.}~\bibnamefont {Buchleitner}},\ }\href@noop {}
  {\bibfield  {journal} {\bibinfo  {journal} {Phys. Rev. Lett.}\ }\textbf
  {\bibinfo {volume} {104}},\ \bibinfo {pages} {220405} (\bibinfo {year}
  {2010})}\BibitemShut {NoStop}%
\bibitem [{\citenamefont {Takesue}\ \emph {et~al.}(2015)\citenamefont
  {Takesue}, \citenamefont {Dyer}, \citenamefont {Stevens}, \citenamefont
  {Verma}, \citenamefont {Mirin},\ and\ \citenamefont {Nam}}]{Takesue2015}%
  \BibitemOpen
  \bibfield  {author} {\bibinfo {author} {\bibfnamefont {H.}~\bibnamefont
  {Takesue}}, \bibinfo {author} {\bibfnamefont {S.~D.}\ \bibnamefont {Dyer}},
  \bibinfo {author} {\bibfnamefont {M.~J.}\ \bibnamefont {Stevens}}, \bibinfo
  {author} {\bibfnamefont {V.}~\bibnamefont {Verma}}, \bibinfo {author}
  {\bibfnamefont {R.~P.}\ \bibnamefont {Mirin}}, \ and\ \bibinfo {author}
  {\bibfnamefont {S.~W.}\ \bibnamefont {Nam}},\ }\href@noop {} {\bibfield
  {journal} {\bibinfo  {journal} {Optica}\ }\textbf {\bibinfo {volume} {2}},\
  \bibinfo {pages} {832} (\bibinfo {year} {2015})}\BibitemShut {NoStop}%
\end{thebibliography}%

\end{document}